\documentclass[twocolumn,journal]{IEEEtran}
\usepackage[T1]{fontenc}
\usepackage{array}
\usepackage{verbatim}
\usepackage{float}
\usepackage{multirow}
\usepackage{amsthm}
\usepackage{amsmath}
\usepackage{amssymb}
\usepackage{graphicx}
\usepackage[unicode=true,
 bookmarks=true,bookmarksnumbered=true,bookmarksopen=true,bookmarksopenlevel=1,
 breaklinks=false,pdfborder={0 0 0},backref=false,colorlinks=false]
 {hyperref}
\hypersetup{pdftitle={Your Title},
 pdfauthor={Your Name},
 pdfpagelayout=OneColumn, pdfnewwindow=true, pdfstartview=XYZ, plainpages=false}

\makeatletter

\providecommand{\tabularnewline}{\\}
\floatstyle{ruled}
\newfloat{algorithm}{tbp}{loa}
\providecommand{\algorithmname}{Algorithm}
\floatname{algorithm}{\protect\algorithmname}

 \let\oldforeign@language\foreign@language
 \DeclareRobustCommand{\foreign@language}[1]{%
   \lowercase{\oldforeign@language{#1}}}
\theoremstyle{plain}
\newtheorem{thm}{\protect\theoremname}
\theoremstyle{remark}
\newtheorem{rem}[thm]{\protect\remarkname}
\theoremstyle{plain}
\newtheorem{lem}[thm]{\protect\lemmaname}

\usepackage[caption=false,font=footnotesize]{subfig}
\usepackage {algorithmic}
\usepackage{cite}

\usepackage{graphicx}
\usepackage{epsfig}
\usepackage{epstopdf}

\makeatother

\providecommand{\lemmaname}{Lemma}
\providecommand{\remarkname}{Remark}
\providecommand{\theoremname}{Theorem}

\begin{document}

\title{Mean-Reverting Portfolio Design with Budget Constraint}

\author{Ziping~Zhao,~\IEEEmembership{Student Member,~IEEE}, and Daniel~P.~Palomar,~\IEEEmembership{Fellow,~IEEE}\thanks{The authors are with the Department of Electronic and Computer Engineering,
The Hong Kong University of Science and Technology (HKUST), Clear
Water Bay, Kowloon, Hong Kong (e-mail: \protect\href{mailto:ziping.zhao@connect.ust.hk}{ziping.zhao@connect.ust.hk};
\protect\href{mailto:palomar@ust.hk}{palomar@ust.hk}).}\thanks{Part of the results in this paper were preliminary presented at \cite{ZhaoPalomar2016}.}}

\markboth{S\MakeLowercase{ubmitted paper}}{ZHAO AND PALOMAR: MRP Design with Budget Constraint}

\maketitle

\begin{abstract}
This paper considers the mean-reverting portfolio design problem arising
from statistical arbitrage in the financial markets. We first propose
a general problem formulation aimed at finding a portfolio of underlying
component assets by optimizing a mean-reversion criterion characterizing
the mean-reversion strength, taking into consideration the variance
of the portfolio and an investment budget constraint. Then several
specific problems are considered based on the general formulation,
and efficient algorithms are proposed. Numerical results on both synthetic
and market data show that our proposed mean-reverting portfolio design
methods can generate consistent profits and outperform the traditional
design methods and the benchmark methods in the literature.\end{abstract}


\IEEEpeerreviewmaketitle{}

\section{Introduction\label{sec:Introduction}}

\IEEEPARstart{P}{airs} trading \cite{Vidyamurthy2004,Ehrman2006,Bookstaber2011,GoetzmannRouwenhorstothers1998,GatevGoetzmannRouwenhorst2006}
is a well-known trading strategy that was pioneered by scientists
Gerry Bamberger and David Shaw, and the quantitative trading group
led by Nunzio Tartaglia at Morgan Stanley in the mid 1980s. As indicated
by the name, it is an investment strategy that focuses on a pair of
assets at the same time. Investors or arbitrageurs embracing this
strategy do not need to forecast the absolute price of every single
asset in one pair, which by nature is hard to assess, but only the
relative price of this pair. As a contrarian investment strategy,
in order to arbitrage from the market, investors need to buy the under-priced
asset and short-sell the over-priced one. Then profits are locked
in after trading positions are unwound when the relative mispricing
corrects itself in the future. 

More generally, pairs trading with only two trading assets falls into
the umbrella of statistical arbitrage \cite{Pole2011,LeRoyWerner2014},
where the underlying trading basket in general consists of three or
more assets. Since profits from such arbitrage strategies do not depend
on the movements and conditions of the general financial markets,
statistical arbitrage is referred to as a kind of market neutral strategies
\cite{JacobsLevy2005,Nicholas2000}. Nowadays, statistical arbitrage
is widely used by institutional investors, hedge fund companies, and
many individual investors in the financial markets.%

In \cite{Granger1983,EngleGranger1987}, the authors first came up
with the concept of cointegration to describe the linear stationary
and hence mean-reverting relationship of underlying nonstationary
time series which are named to be cointegrated. Later, the cointegrated
vector autoregressive model \cite{Johansen1988,Johansen1991,LarssonJohansen1997,Johansen2000,Johansen1995}
was proposed to describe the cointegration relations. Empirical and
technical analyses \cite{AvellanedaLee2010,DunisGiorgioniLawsEtAl2010,CaldeiraMoura2012,Drakos2016}
show that cointegration can be used to get statistical arbitrage opportunities
and such relations really exist in financial markets. Taking the prices
of common stocks for example, it is generally known that a stock price
is observed and modeled as a nonstationary random walk process that
can be hard to predict efficiently. However, companies in the same
financial sector or industry usually share similar fundamental characteristics,
then their stock prices may move in company with each other under
the same trend, based on which cointegration relations can be established.
Two examples are the stock prices of the two American famous consumer
staple companies Coca-Cola and PepsiCo and those of the two energy
companies Ensco and Noble Corporation. Some examples for other financial
assets, to name a few, are the future contract prices of E-mini S\&P
500 and E-mini Dow, the ETF prices of SPDR S\&P 500 and SPDR DJIA,
the US dollar foreign exchange rates for different countries, and
the swap rates for US interest rates of different maturities.%

Mean-reversion is a classic indicator of predictability in financial
markets and used to obtain arbitrage opportunities. Assets in a cointegration
relation can be used to form a portfolio or basket and traded based
upon their stationary mean-reversion property. We call such a designed
portfolio or basket of underlying assets a mean-reverting portfolio
(MRP) or sometimes a long-short portfolio which is also called a ``spread''.
An asset whose price shows naturally stationarity is a spread as well.
The profits of statistical arbitrage come directly from trading on
the mean-reversion of the spread around the long-run equilibrium.
MRPs in practice are usually constructed using heuristic or statistical
methods. Traditional statistical cointegration estimation methods
are Engle-Granger ordinary least squares (OLS) method \cite{EngleGranger1987}
and Johansen model-based method \cite{Johansen1991}. In practice,
inherent correlations may exist among different MRPs. However, when
having multiple MRPs, they are commonly traded separately with their
possible connections neglected. So a natural and interesting question
is whether we can design an optimized MRP based on the underlying
spreads that could outperform every single one. In this paper, this
issue is clearly addressed.%

Designing one MRP by choosing proportions of various assets in general
is a portfolio optimization or asset allocation problem \cite{FarrellReinhart1997}.
Portfolio optimization today is considered to be an important part
in portfolio management as well as in algorithmic trading. The seminal
paper \cite{Markowitz1952} by Markowitz in 1952 laid on the foundations
of what is now popularly referred to as mean-variance portfolio optimization
and modern portfolio theory. Given a collection of financial assets,
the traditional mean-variance portfolio design problem is aimed at
finding a tradeoff between the expected return and the risk measured
by the variance. Different from the requirements for mean-variance
portfolio design, in order to design a mean-reverting portfolio, there
are two main factors to consider: i) the designed MRP should exhibit
a strong mean-reversion indicating that it should have frequent mean-crossing
points and hence bring in trading opportunities, and ii) the designed
MRP should exhibit sufficient but controlled variance so that each
trade can provide enough profit while controlling the probability
that the believed mean-reversion equilibrium breaks down could be
reduced.%

In \cite{dAspremont2011}, the author first proposed to design an
MRP by optimizing a criterion characterizing the mean-reversion strength
which is a model-free method. Later, authors in \cite{CuturidAspremont2013}
realized that solving the problem in \cite{dAspremont2011} could
result in a portfolio with very low variance, then the variance control
was taken into consideration and also new criteria to characterize
the mean-reversion property were proposed for the MRP design problem.
In \cite{dAspremont2011,CuturidAspremont2013}, semidefinite programming
(SDP) relaxation methods were used to solve the nonconvex problem
formulations; however, these methods are very computationally costly
in general. Besides that, the design methods in \cite{dAspremont2011,CuturidAspremont2013}
were all carried out by imposing an $\ell_{2}$-norm constraint on
the portfolio weights. This constraint brings mathematically convenience
to the optimization problem, but its practical significance in financial
applications is dubious since the $\ell_{2}$-norm is not meaningful
in a financial context. In this paper, we propose to use investment
budget constraints in the design problems. %
{} 

The contributions of this paper can be summarized as follows. First,
a general problem formulation for MRP design problem is proposed based
on which several specific problem formulations are elaborated by considering
different mean-reversion criteria. Second, Two classes of commonly
used investment budget constraints on portfolio weights are considered,
namely, dollar neutral constraint and net budget constraint. Third,
efficient algorithms are proposed for the proposed problem formulations,
it is shown that some problems after reformulations can be tackled
readily by solving the well-known generalized eigenvalue problem (GEVP)
and the generalized trust region subproblem (GTRS). The other problems
can be easily solved based on the majorization-minimization (MM) framework
by solving a sequence of GEVPs and GTRSs, which are named iteratively
reweighted generalized eigenvalue problem (IRGEVP) and iteratively
reweighted generalized trust region subproblem (IRGTRS), respectively.
An extension for IRGEVP with closed-form solution in every iteration
named EIRGEVP (extended IRGEVP) is also proposed. %

The remaining sections of this paper are organized as follows. In
Section \ref{sec:Mean-Reverting-Portfolio}, we introduce the design
of mean-reverting portfolios. In Section \ref{sec:MRP-Design-Problem-Formulations},
we give out some mean-reversion criteria for an MRP, and a general
formulation for the MRP design problem is proposed together with two
commonly used investment budget constraint. Section \ref{sec:Problem-Solving-Methods-GEVP-GTRS}
develops the solving methods for GEVP and GTRS. The MM framework and
MM-based solving algorithms are elaborated in Section \ref{sec:Problem-Solving-Methods-MM}.
The performance of the proposed algorithms are evaluated numerically
in Section \ref{sec:Numerical-Experiments} and, finally, the concluding
remarks are drawn in Section \ref{sec:Conclusions}.%

\textsl{Notation}\textit{:} Boldface upper case letters denote matrices,
boldface lower case letters denote column vectors, and italics denote
scalars. The notation $\mathbf{1}$ and $\mathbf{I}$ denote an all-one
column vector and an identity matrix with proper size, respectively.
$\mathbb{R}$ denotes the real field with $\mathbb{R}^{+}$ denoting
positive real numbers and $\mathbb{R}^{N}$ denoting the $N$-dimensional
real vector space. $\mathbb{N}$ denotes the natural field. $\mathbb{Z}$
denotes the integer circle with $\mathbb{Z}^{+}$ denoting positive
integer numbers. $\mathbb{S}^{K}$ denotes the $K\times K$-dimensional
symmetric matrices. The superscripts $\left(\cdot\right)^{T}$ and
$\left(\cdot\right)^{-1}$ denote the matrix transpose and inverse
operator, respectively. Due to the commutation of the inverse and
the transpose for nonsingular matrices, the superscript $\left(\cdot\right)^{-T}$
denotes the matrix inverse and transpose operator. $x_{i,j}$ denotes
the ($i$th, \textbf{$j$}th) element of matrix $\mathbf{X}$ and
$x_{i}$ denotes the $i$th element of vector $\mathbf{x}$. $\text{Tr}\left(\cdot\right)$
denotes the trace of a matrix. $\text{vec}\left(\cdot\right)$ denotes
the vectorization of a matrix, i.e., $\text{vec}\left(\mathbf{X}\right)$
is a column vector consisting of all the columns of $\mathbf{X}$
stacked. $\otimes$ denotes the Kronecker product of two matrices.

\section{Mean-Reverting Portfolio (MRP)\label{sec:Mean-Reverting-Portfolio}}

For a financial asset, e.g., a common stock, a future contract, an
ETF, or a portfolio of them, its price at time index or holding period
$t\in\mathbb{Z}^{+}$ is denoted by $p_{t}\in\mathbb{R}^{+}$, and
the corresponding logarithmic price or log-price $y_{t}\in\mathbb{R}$
is computed as $y_{t}=\log\left(p_{t}\right)$, where $\log\left(\cdot\right)$
is the natural logarithm. 

If we consider a collection of $M$ assets in a basket, their log-prices
can be accordingly denoted by $\mathbf{y}_{t}=\left[y_{1,t},y_{2,t},\ldots,y_{M,t}\right]^{T}\in\mathbb{R}^{M}$.
Based on this basket, an MRP is accordingly defined by the portfolio
weight or hedge ratio $\mathbf{w}_{s}=\left[w_{s,1},w_{s,2},\ldots,w_{s,M}\right]^{T}\in\mathbb{R}^{M}$
and its (log-price) spread $s_{t}$ is defined as $s_{t}=\mathbf{w}_{s}^{T}\mathbf{y}_{t}=\sum_{m=1}^{M}w_{s,m}y_{m,t}$.
Vector $\mathbf{w}_{s}$ indicates the market value proportion invested
on the underlying asset\footnote{If the spread is designed based on asset price $p_{t}$ instead of
the log-price, $\mathbf{w}_{s}$ indicates the asset amount proportion
measured in shares.}. For $m=1,2,\ldots,M$, $w_{s,m}>0$, $w_{s,m}<0$, and $w_{s,m}=0$
mean a long position (i.e., the asset is bought), a short position
(i.e., the asset is short-sold, or, more plainly, borrowed and sold),
and no position, respectively. 

\begin{figure}[h]
\centering{}\includegraphics[scale=0.63]{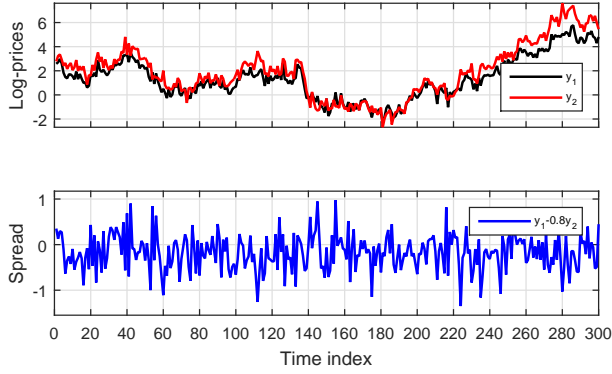}\protect\caption{\label{fig:An-example}An illustrative example of log-prices of two
assets and a designed spread.}
\end{figure}

In Figure \ref{fig:An-example}, the spread of a designed MRP together
with the log-prices of the two underlying assets is given. It is worth
noting that an MRP can be interpreted as a synthesized stationary
asset. The spread accordingly means its log-price which could be easier
to predict and to make profits from in comparison with the underlying
component assets in this MRP. 

Suppose there exist $N$ MRPs with their spreads denoted by $\mathbf{s}_{t}=\left[s_{1,t},s_{2,t},\ldots,s_{N,t}\right]^{T}\in\mathbb{R}^{N}$.
Different spreads may possess different mean-reversion and variance
properties in nature. Our objective is to design an MRP to combine
such spreads into an improved overall spread with better properties.
In particular, we denote the portfolio by $\mathbf{w}=\left[w_{1},w_{2},\ldots,w_{N}\right]^{T}\in\mathbb{R}^{N}$,
where $\mathbf{w}$ denotes the market value on the underlying spread.
The resulting overall spread is then given by

\begin{equation}
z_{t}=\mathbf{w}^{T}\mathbf{s}_{t}=\sum_{n=1}^{N}w_{n}s_{n,t}.\label{eq:spread for MRP}
\end{equation}

\section{MRP Design Problem Formulations\label{sec:MRP-Design-Problem-Formulations}}

Traditional portfolio design problems are based on the Nobel prize-winning
Markowitz portfolio theory \cite{Markowitz1952,Markowitz1956,Sharpe1964,Markowitz1968}.
They aim at finding a desired trade-off between return and risk, the
latter being measured traditionally by the variance or, in a more
sophisticated way, by value-at-risk and conditional value-at-risk.
The recently proposed risk-parity portfolios \cite{Qian2011,ChavesHsuLiShakernia2011,FengPalomar2015}
can also be categorized into this design problem. 

For the mean-reverting portfolio, we can formulate the design problem
by optimizing some mean-reversion criterion quantifying the mean-reversion
strength of the spread $z_{t}$, while controlling its variance and
imposing an investment budget constraint.%

\subsection{Mean-Reversion Criteria\label{sub:Mean-Reversion-Criteria}}

In this section, we introduce several mean-reversion criteria that
can characterize the mean-reversion strength of the designed spread
$z_{t}$.%
{} We start by defining the $i$th order (lag-$i$) autocovariance matrix
for a stochastic process $\mathbf{s}_{t}$ as

\begin{equation}
\begin{array}{ccl}
\mathbf{M}_{i} & = & \mathsf{Cov}\left(\mathbf{s}_{t},\mathbf{s}_{t+i}\right)\\
 & = & \mathsf{E}\left[\left(\mathbf{s}_{t}-\mathsf{E}\left[\mathbf{s}_{t}\right]\right)\left(\mathbf{s}_{t+i}-\mathsf{E}\left[\mathbf{s}_{t+i}\right]\right)^{T}\right],
\end{array}\label{eq:autocovariance for s}
\end{equation}
where $i\in\mathbb{N}$. Specifically, when $i=0$, $\mathbf{M}_{0}$
stands for the (positive definite) covariance matrix of $\mathbf{y}_{t}$. 

Since for any random process $\mathbf{s}_{t}$, we can get its centered
counterpart $\tilde{\mathbf{s}}_{t}$ as $\tilde{\mathbf{s}}_{t}=\mathbf{s}_{t}-\mathsf{E}\left[\mathbf{s}_{t}\right]$,
in the following, we will use $\mathbf{s}_{t}$ to denote its centered
form $\tilde{\mathbf{s}}_{t}$. %

\subsubsection{Predictability Statistics $\mathrm{pre}_{z}\left(\mathbf{w}\right)$\label{sub:Predictability-Statistics}}

Consider a centered univariate stationary autoregressive process written
as follows:
\begin{equation}
z_{t}=\hat{z}_{t-1}+\epsilon_{t},
\end{equation}
where $\hat{z}_{t-1}$ is the prediction of $z_{t}$ based on the
information up to time $t-1$, and $\epsilon_{t}$ denotes a white
noise independent from $\hat{z}_{t-1}$. The predictability statistics
\cite{BoxTiao1977} is defined as

\begin{equation}
\mathrm{pre}_{z}=\frac{\sigma_{\hat{z}}^{2}}{\sigma_{z}^{2}},\label{eq:predictability statistics}
\end{equation}
where $\sigma_{z}^{2}=\mathsf{E}\left[z_{t}^{2}\right]$ and $\sigma_{\hat{z}}^{2}=\mathsf{E}\left[\hat{z}_{t-1}^{2}\right]$.
If we define $\sigma_{\epsilon}^{2}=\mathsf{E}\left[\epsilon_{t}^{2}\right]$,
then from \eqref{eq:predictability statistics}, we can have $\sigma_{z}^{2}=\sigma_{\hat{z}}^{2}+\sigma_{\epsilon}^{2}$
in the denominator. When $\mathrm{pre}_{z}$ is small, the variance
of $\epsilon_{t}$ dominates that of $\hat{z}_{t-1}$, and $z_{t}$
behaves like a white noise; when $\mathrm{pre}_{z}$ is large, the
variance of $\hat{z}_{t-1}$ dominates that of $\epsilon_{t}$, and
$z_{t}$ can be well predicted by $\hat{z}_{t-1}$. The predictability
statistics is usually used to measure how close a random process is
to a white noise. 

Under this criterion, in order to design a spread $z_{t}$ as close
as possible to a white noise process, we need to minimize $\mathrm{pre}_{z}$.
For a spread $z_{t}=\mathbf{w}^{T}\mathbf{s}_{t}$, we assume the
spread $\mathbf{s}_{t}$ follows a centered vector autoregressive
model of order $1$ (VAR($1$)) as follows:
\begin{equation}
\mathbf{s}_{t}=\mathbf{A}\mathbf{s}_{t-1}+\mathbf{e}_{t},\label{eq:VAR(1)}
\end{equation}
where $\mathbf{A}$ is the autoregressive coefficient and $\mathbf{e}_{t}$
denotes a white noise independent from $\mathbf{s}_{t-1}$.%
{} We can get $\mathbf{A}$ from the autocorrelation matrices as $\mathbf{A}=\mathbf{M}_{1}\mathbf{M}_{0}^{-1}$.
Multiplying \eqref{eq:VAR(1)} by $\mathbf{w}$ and further defining
$\hat{z}_{t-1}=\mathbf{w}^{T}\mathbf{A}\mathbf{s}_{t-1}$ and $\epsilon_{t}=\mathbf{w}^{T}\mathbf{e}_{t}$,
we can get $\sigma_{z}^{2}=\mathbf{w}^{T}\mathbf{M}_{0}\mathbf{w}$,
and $\sigma_{\hat{z}}^{2}=\mathbf{w}^{T}\mathbf{T}\mathbf{w}$, where
$\mathbf{T}=\mathbf{A}\mathbf{M}_{0}\mathbf{A}^{T}=\mathbf{M}_{1}\mathbf{M}_{0}^{-1}\mathbf{M}_{1}^{T}$.%
{} High order models VAR($p$), with $p>1$, can be trivially reformulated
into VAR($1$) with proper reparametrization \cite{Luetkepohl2007}.
Then the estimator of predictability statistics for $z_{t}$ is computed
as
\begin{equation}
\mathrm{pre}_{z}\left(\mathbf{w}\right)=\frac{\mathbf{w}^{T}\mathbf{T}\mathbf{w}}{\mathbf{w}^{T}\mathbf{M}_{0}\mathbf{w}}.\label{eq:predictability statistics using w}
\end{equation}

\subsubsection{Portmanteau Statistics $\mathrm{por}_{z}\left(p,\mathbf{w}\right)$\label{sub:Portmanteau-Statistics}}

The portmanteau statistics of order $p$ \cite{BoxPierce1970} for
a centered univariate stationary process $z_{t}$ is defined as
\begin{equation}
\mathrm{por}_{z}\left(p\right)=T\sum_{i=1}^{p}\rho_{i}^{2},\label{eq:portmanteau statistics}
\end{equation}
where $\rho_{i}$ is the $i$th order autocorrelation (autocorrelation
for lag $i$) of $z_{t}$ defined as $\rho_{i}=\frac{\mathsf{E}\left[z_{t}z_{t+i}\right]}{\mathsf{E}\left[z_{t}^{2}\right]}$.
The portmanteau statistics is used to test whether a random process
is close to a white noise. From the above definition, we have $\mathrm{por}_{z}\left(p\right)\geq0$
and the minimum of $\mathrm{por}_{z}\left(p\right)$ is attained by
a white noise process, i.e., the portmanteau statistics for a white
noise process is \textbf{$0$} for any $p$. 

Under this criterion, in order to get a spread $z_{t}$ close to a
white noise process, we need to minimize $\text{por}_{z}\left(p\right)$
for a prespecified order $p$. For an MRP $z_{t}=\mathbf{w}^{T}\mathbf{s}_{t}$,
the $i$th order autocorrelation is given by 
\begin{equation}
\rho_{i}=\frac{\mathsf{E}\left[z_{t}z_{t+i}\right]}{\mathsf{E}\left[z_{t}^{2}\right]}=\frac{\mathbf{w}^{T}\mathsf{E}\left[\mathbf{s}_{t}\mathbf{s}_{t+i}^{T}\right]\mathbf{w}}{\mathbf{w}^{T}\mathsf{E}\left[\mathbf{s}_{t}\mathbf{s}_{t}^{T}\right]\mathbf{w}}=\frac{\mathbf{w}^{T}\mathbf{M}_{i}\mathbf{w}}{\mathbf{w}^{T}\mathbf{M}_{0}\mathbf{w}}.\label{eq:autocorrelation for z}
\end{equation}
Then we can get the expression for $\mathrm{por}_{z}\left(p,\mathbf{w}\right)$
as
\begin{equation}
\mathrm{por}_{z}\left(p,\mathbf{w}\right)=T\sum_{i=1}^{p}\left(\frac{\mathbf{w}^{T}\mathbf{M}_{i}\mathbf{w}}{\mathbf{w}^{T}\mathbf{M}_{0}\mathbf{w}}\right)^{2}.\label{eq:portmanteau statistics using w}
\end{equation}

\subsubsection{Crossing Statistics $\mathrm{cro}_{z}\left(\mathbf{w}\right)$ and
Penalized Crossing Statistics $\mathrm{pcro}_{z}\left(p,\mathbf{w}\right)$\label{sub:Crossing-Statistics}}

Crossing statistics (zero-crossing rate) of a centered univariate
stationary process $z_{t}$ is defined as
\begin{equation}
\mathrm{cro}_{z}=\frac{1}{T-1}\mathsf{E}\left[\sum_{t=2}^{T}{\bf 1}_{\left\{ z_{t}z_{t-1}\leq0\right\} }\right],\label{eq:crossing statistics}
\end{equation}
where ${\bf 1}_{E}\left(z_{t}\right)$ is the indicator function defined
as ${\bf 1}_{E}\left(z_{t}\right)=\begin{cases}
1, & \text{if }z_{t}\in E\\
0, & \text{if }z_{t}\notin E
\end{cases}$ , and the event here is $E=\left\{ z_{t}z_{t-1}\leq0\right\} $.
Crossing statistics is used to test the probability that a stationary
process crosses its mean per unit of time and it is easy to notice
that $\mathrm{cro}_{z}\in\left[0,1\right]$. According to \cite{Ylvisaker1965,KedemYakowitz1994},
for a centered stationary Gaussian process $z_{t}$, we have the following
relationship:

\begin{equation}
\mathrm{cro}_{z}=\frac{1}{\pi}\arccos\left(\rho_{1}\right).
\end{equation}

\begin{rem}
As a special case, if $z_{t}$ is a centered stationary AR(1),
\begin{equation}
z_{t}=\phi z_{t-1}+\epsilon_{t},\label{eq:AR(1)}
\end{equation}
where $\left|\phi\right|<1$ and $\epsilon_{t}$ is a Gaussian white
noise, then $\phi=\rho_{1}$ and accordingly the crossing statistics
is $\mathrm{cro}_{z}=\frac{1}{\pi}\arccos\left(\phi\right)$.
\end{rem}
Using this criterion, in order to get a spread $z_{t}$ having many
zero-crossings, instead of directly maximizing $\mathrm{cro}_{z}$,
we can minimize $\rho_{1}$. For a spread $z_{t}=\mathbf{w}^{T}\mathbf{s}_{t}$,
we can try to minimize the first order autocorrelation of $z_{t}$
given in \eqref{eq:autocorrelation for z}. In \cite{CuturidAspremont2013},
besides minimizing the first order autocorrelation, it is also proposed
to ensure the absolute autocorrelations of high orders $\left|\rho_{i}\right|$s
($i=2,\ldots,p$) are small at the same time which can result in good
performance. In this paper, we also adopt this criterion and call
it penalized crossing statistics of order $p$ defined by

\begin{equation}
\mathrm{pcro}_{z}\left(p,\mathbf{w}\right)=\frac{\mathbf{w}^{T}\mathbf{M}_{1}\mathbf{w}}{\mathbf{w}^{T}\mathbf{M}_{0}\mathbf{w}}+\eta\sum_{i=2}^{p}\left(\frac{\mathbf{w}^{T}\mathbf{M}_{i}\mathbf{w}}{\mathbf{w}^{T}\mathbf{M}_{0}\mathbf{w}}\right)^{2},
\end{equation}
where $\eta$ is a positive prespecified penalization factor.

\subsection{General MRP Design Problem Formulation\label{sub:General-Mean-Reversion-Portfolio-Formulation}}

The MRP design problem is formulated as the optimization of a mean-reversion
criterion denoted in general as $F_{z}\left(\mathbf{w}\right)$, which
can be taken to be any of the criteria mentioned before. This unified
criterion can be written into a compact form as 
\begin{equation}
\begin{array}{c}
F_{z}\left(\mathbf{w}\right)=\xi\frac{\mathbf{w}^{T}\mathbf{H}\mathbf{w}}{\mathbf{w}^{T}\mathbf{M}_{0}\mathbf{w}}+\zeta\left(\frac{\mathbf{w}^{T}\mathbf{M}_{1}\mathbf{w}}{\mathbf{w}^{T}\mathbf{M}_{0}\mathbf{w}}\right)^{2}+\eta\sum_{i=2}^{p}\left(\frac{\mathbf{w}^{T}\mathbf{M}_{i}\mathbf{w}}{\mathbf{w}^{T}\mathbf{M}_{0}\mathbf{w}}\right)^{2},\end{array}\label{eq:objective of MRP design problem formulation}
\end{equation}
which particularizes to i) $\mathrm{pre}_{z}\left(\mathbf{w}\right)$,
when $\xi=1$, $\mathbf{H}=\mathbf{T}$, and $\zeta=\eta=0$; ii)
$\mathrm{por}_{z}\left(p,\mathbf{w}\right)$, when $\xi=0$, and $\zeta=\eta=1$;
iii) $\mathrm{cro}_{z}\left(\mathbf{w}\right)$, when $\xi=1$, $\mathbf{H}=\mathbf{M}_{1}$,
and $\zeta=\eta=0$; and iv) $\mathrm{pcro}_{z}\left(p,\mathbf{w}\right)$,
when $\xi=1$, $\mathbf{H}=\mathbf{M}_{1}$, $\zeta=0$, and $\eta>0$.
The matrices $\mathbf{M}_{i}$s in \eqref{eq:objective of MRP design problem formulation}
are assumed symmetric without loss of generality since they can be
symmetrized. 

The variance of the spread should also be controlled to a certain
level which can be represented as $\mathsf{Var}\left[\mathbf{w}^{T}\mathbf{s}_{t}\right]=\mathbf{w}^{T}\mathbf{M}_{0}\mathbf{w}=\nu$.
Due to this variance constraint, the denominators of $F_{z}\left(\mathbf{w}\right)$
can be removed. Denoting the portfolio investment budget constraint
by ${\cal W}$, the general MRP design problem can be formulated as
follows:

\begin{equation}
\begin{array}{ll}
\underset{\mathbf{w}}{\mathsf{mininize}} & \xi\mathbf{w}^{T}\mathbf{H}\mathbf{w}+\zeta\left(\mathbf{w}^{T}\mathbf{M}_{1}\mathbf{w}\right)^{2}+\eta\sum_{i=2}^{p}\left(\mathbf{w}^{T}\mathbf{M}_{i}\mathbf{w}\right)^{2}\\
\mathsf{subject\:to} & \mathbf{w}^{T}\mathbf{M}_{0}\mathbf{w}=\nu\\
 & \mathbf{w}\in{\cal W},
\end{array}\label{eq:MRP design problem formulation}
\end{equation}
where the objective function is denoted by $f_{z}\left(\mathbf{w}\right)$
in the following. The problem in \eqref{eq:MRP design problem formulation}
is a nonconvex problem due to the nonconvexity of the objective function
and the constraint set.

\subsection{Investment Budget Constraint ${\cal W}$\label{sub:Portfolio-Weight-Constraint}}

In portfolio optimization, constraints are usually imposed to represent
the specific investment guidelines. In this paper, we use ${\cal W}$
to denote it and we focus on two types of budget constraints: dollar
neutral constraint and net budget constraint. 

Dollar neutral constraint, denoted by ${\cal W}_{0}$, means the net
investment or net portfolio position is zero; in other words, all
the long positions are financed by the short positions, commonly termed
self-financing.\footnote{Dollar neutral constraint generally cannot be satisfied by the traditional
design methods, like methods in \cite{EngleGranger1987} and \cite{Johansen1991},
and the methods in \cite{CuturidAspremont2013}.} It is represented mathematically by 
\begin{equation}
{\cal W}_{0}=\left\{ \ensuremath{\mathbf{1}^{T}\mathbf{w}=0}\right\} .\label{eq:dollar neutral constraint}
\end{equation}

Net budget constraint, denoted by ${\cal W}_{1}$, means the net investment
or net portfolio position is nonzero and equal to the current budget
which is normalized to one.\footnote{The net portfolio position can be positive or negative under net budget
constraint. Since the problem formulation in \eqref{eq:MRP design problem formulation}
is invariant to the sign of $\mathbf{w}$, only the case that budget
is normalized to positive 1 is considered.} It is represented mathematically by%
{} 
\begin{equation}
{\cal W}_{1}=\left\{ \ensuremath{\mathbf{1}^{T}\mathbf{w}=1}\right\} .\label{eq:net budget constraint}
\end{equation}

It is worth noting that, for two trading spreads defined by $\mathbf{w}^{T}\mathbf{y}_{t}$
and $-\mathbf{w}^{T}\mathbf{y}_{t}$, they are naturally the same,
because in statistical arbitrage the actual investment not only depends
on $\mathbf{w}$, which defines a spread, but also on whether a long
or short position is taken on this spread in the trading.%

\section{Problem Solving Algorithms via GEVP and GTRS Algorithms\label{sec:Problem-Solving-Methods-GEVP-GTRS}}

In this section, solving methods for the MRP design problem formulations
using $\mathrm{pre}_{z}\left(\mathbf{w}\right)$ and $\mathrm{cro}_{z}\left(\mathbf{w}\right)$
(i.e., \eqref{eq:MRP design problem formulation} with $\zeta=\eta=0$)
are introduced.

\subsection{GEVP - Solving Algorithm for MRP Design Using $\mathrm{pre}_{z}\left(\mathbf{w}\right)$
and $\mathrm{cro}_{z}\left(\mathbf{w}\right)$ with $\mathbf{w}\in\mathcal{W}_{0}$\label{sub:GEVP}}

For notational simplicity, we denote the matrices $\mathbf{T}$ in
$\mathrm{pre}_{z}\left(\mathbf{w}\right)$ and $\mathbf{M}_{1}$ in
$\mathrm{cro}_{z}\left(\mathbf{w}\right)$ by matrix $\mathbf{H}$
in general and recast the problem as follows: 

\begin{equation}
\begin{array}{ll}
\underset{\mathbf{w}}{\mathsf{minimize}} & \mathbf{w}^{T}\mathbf{H}\mathbf{w}\\
\mathsf{subject\:to} & \mathbf{w}^{T}\mathbf{M}_{0}\mathbf{w}=\nu\\
 & \mathbf{1}^{T}\mathbf{w}=0,
\end{array}\label{eq:pre and cro statistics with eq0}
\end{equation}
where $\nu$ is a positive constant. The above problem is equivalent
to the following nonconvex quadratically constrained quadratic programming
(QCQP) \cite{BoydVandenberghe2004} formulation:

\begin{equation}
\begin{array}{ll}
\underset{\mathbf{w}}{\mathsf{minimize}} & \mathbf{w}^{T}\mathbf{H}\mathbf{w}\\
\mathsf{subject\:to} & \mathbf{w}^{T}\mathbf{M}_{0}\mathbf{w}=\nu\\
 & \mathbf{w}^{T}\mathbf{E}\mathbf{w}=0,
\end{array}\label{eq:pre and cro statistics with eq0-QCQP}
\end{equation}
where $\mathbf{E}=\mathbf{1}\mathbf{1}^{T}$. By using the matrix
lifting technique, i.e., defining $\mathbf{W}=\mathbf{w}\mathbf{w}^{T}$,
the above problem can be solved by the following convex SDP relaxation
problem:

\begin{equation}
\begin{array}{ll}
\underset{\mathbf{W}}{\mathsf{minimize}} & \text{Tr}\left(\mathbf{H}\mathbf{W}\right)\\
\mathsf{subject\:to} & \text{Tr}\left(\mathbf{M}_{0}\mathbf{W}\right)=\nu\\
 & \text{Tr}\left(\mathbf{E}\mathbf{W}\right)=0\\
 & \mathbf{W}\succeq\mathbf{0}.
\end{array}\label{eq:pre and cro statistics with eq0-SDP}
\end{equation}
The following theorem gives a useful relationship between the number
of variables and the number of equality constraints.
\begin{thm}[{\cite[Theorem 3.2]{HuangPalomar2010}}]
\label{thm:HuangPalomar} Given a separable SDP as follows:
\begin{equation}
\begin{array}{ll}
\underset{\mathbf{X}_{1},\ldots,\mathbf{X}_{L}}{\mathsf{minimize}} & \sum_{l=1}^{L}\text{Tr}\left(\mathbf{A}_{l}\mathbf{X}_{l}\right)\\
\mathsf{subject\:to} & \sum_{l=1}^{L}\text{Tr}\left(\mathbf{B}_{ml}\mathbf{X}_{l}\right)=b_{m},\;m=1,\ldots,M\\
 & \mathbf{X}_{l}\succeq\mathbf{0},\;l=1,\ldots,L.
\end{array}\label{eq:SDP}
\end{equation}
Suppose that the separable SDP are strictly feasible. Then, the problem
has always an optimal solution $\left(\mathbf{X}_{1}^{\star},\ldots,\mathbf{X}_{L}^{\star}\right)$
such that

\[
\sum_{l=1}^{L}\left[\text{rank}\left(\mathbf{X}{}_{l}^{\star}\right)\right]^{2}\leq M.
\]

\end{thm}
Observe that if there is only one variable $\mathbf{X}$,\textbf{
}that is to say,\textbf{ $L=1$}, we can get $\text{rank}\left(\mathbf{X}^{\star}\right)\leq\sqrt{M}$.
Further, if the number of constraints $M\leq3$, a rank-$1$ solution
can always be attainable.%

\begin{lem}
\label{lem:QCQP eq0}The nonconvex problem in \eqref{eq:pre and cro statistics with eq0}
or \eqref{eq:pre and cro statistics with eq0-QCQP} has no duality
gap.\end{lem}
\begin{IEEEproof}
This lemma directly follows from Theorem \ref{thm:HuangPalomar} and
the equivalence of problems \eqref{eq:pre and cro statistics with eq0}
and \eqref{eq:pre and cro statistics with eq0-QCQP}.
\end{IEEEproof}
In other words, by solving the convex SDP in \eqref{eq:pre and cro statistics with eq0-SDP},
there always exists a rank\textbf{-$1$} solution for $\mathbf{W}$
which is the solution for the original problem \eqref{eq:pre and cro statistics with eq0},
however, in practice, to find such a solution, rank reduction methods
\cite{HuangPalomar2014} should be applied which could be computationally
expensive. 

As an alternative to the SDP procedure mentioned above, we find the
problem in \eqref{eq:pre and cro statistics with eq0} can be efficiently
solved as a generalized eigenvalue problem (GEVP) \cite{HornJohnson2012}
by reformulation. Considering $\mathbf{w}=\mathbf{F}\mathbf{x}$,
where $\mathbf{F}$ is the kernel that spans the null space of $\mathbf{1}^{T}$,
i.e., $\mathbf{1}^{T}\mathbf{F}=\mathbf{0}$, and also required to
be semi-unitary, i.e., $\mathbf{F}^{T}\mathbf{F}=\mathbf{I}$, we
can define $\mathbf{N}=\mathbf{F}^{T}\mathbf{H}\mathbf{F}$ and $\mathbf{N}_{0}=\mathbf{F}^{T}\mathbf{M}_{0}\mathbf{F}$
which is positive definite, then the problem \eqref{eq:pre and cro statistics with eq0}
is equivalent to the following problem:

\begin{equation}
\begin{array}{ll}
\underset{\mathbf{x}}{\mathsf{minimize}} & \mathbf{x}^{T}\mathbf{N}\mathbf{x}\\
\mathsf{subject\:to} & \mathbf{x}^{T}\mathbf{N}_{0}\mathbf{x}=\nu,
\end{array}\label{eq:GEVP}
\end{equation}
which is still a nonconvex QCQP. However, this problem becomes the
classical GEVP problem and can be easily dealt with using tailored
algorithms. Here, we will apply the steepest descent algorithm \cite{SongBabuPalomar2015}
to solve it. The procedure to solve problem \eqref{eq:pre and cro statistics with eq0}
is summarized in Algorithm \ref{alg:GEVP---Algorithm}.%

\begin{algorithm}[tbh]
\begin{algorithmic}[1]

\REQUIRE $\mathbf{N}$, $\mathbf{N}_{0}$, and $\nu>0$.

\STATE Set $k=0$, and choose $\mathbf{x}^{\left(k\right)}\in\left\{ \mathbf{x}:\mathbf{x}^{T}\mathbf{N}_{0}\mathbf{x}=\nu\right\}$;

\REPEAT
\STATE Compute $R\left(\mathbf{x}^{\left(k\right)}\right)=\mathbf{x}^{\left(k\right)T}\mathbf{N}\mathbf{x}^{\left(k\right)}/\mathbf{x}^{\left(k\right)T}\mathbf{N}_{0}\mathbf{x}^{\left(k\right)}$;

\STATE Compute $\mathbf{d}^{\left(k\right)}=\mathbf{N}\mathbf{x}^{\left(k\right)}-R\left(\mathbf{x}^{\left(k\right)}\right)\mathbf{N}_{0}\mathbf{x}^{\left(k\right)}$;

\STATE $\mathbf{x}=\mathbf{x}^{\left(k\right)}+\tau\mathbf{d}^{\left(k\right)}$ with $\tau$ chosen to minimize $R\left(\mathbf{x}^{\left(k\right)}+\tau\mathbf{d}^{\left(k\right)}\right)$;
\STATE $\mathbf{x}^{\left(k+1\right)}=\sqrt{\nu}\mathbf{x}/\sqrt{\mathbf{x}^{T}\mathbf{N}_{0}\mathbf{x}}$;
\STATE $k=k+1$;
\UNTIL convergence

\end{algorithmic} 

\protect\caption{\label{alg:GEVP---Algorithm}GEVP - Algorithm for MRP design problems
using $\mathrm{pre}_{z}\left(\mathbf{w}\right)$ and $\mathrm{cro}_{z}\left(\mathbf{w}\right)$
with $\mathbf{w}\in\mathcal{W}_{0}$.}
\end{algorithm}

\subsection{GTRS - Solving Algorithm for MRP Design Using $\mathrm{pre}_{z}\left(\mathbf{w}\right)$
and $\mathrm{cro}_{z}\left(\mathbf{w}\right)$ with $\mathbf{w}\in\mathcal{W}_{1}$\label{sub:GTRS}}

As before, for generality, we denote matrices $\mathbf{T}$ in $\mathrm{pre}_{z}\left(\mathbf{w}\right)$
and $\mathbf{M}_{1}$ in $\mathrm{cro}_{z}\left(\mathbf{w}\right)$
as $\mathbf{H}$. Then the problems can be rewritten as 

\begin{equation}
\begin{array}{ll}
\underset{\mathbf{w}}{\mathsf{minimize}} & \mathbf{w}^{T}\mathbf{H}\mathbf{w}\\
\mathsf{subject\:to} & \mathbf{w}^{T}\mathbf{M}_{0}\mathbf{w}=\nu\\
 & \mathbf{1}^{T}\mathbf{w}=1,
\end{array}\label{eq:pre and cro statistics with eq1}
\end{equation}
where $\nu$ is a positive constant. As before, rewriting the constraint
$\mathbf{1}^{T}\mathbf{w}=1$ as $\mathbf{w}^{T}\mathbf{E}\mathbf{w}=1$
(since the problem is invariant with respect to a sign change in $\mathbf{w}$)
and using the matrix lifting technique, the problem in \eqref{eq:pre and cro statistics with eq1}
can be solved by the following convex SDP problem:%

\begin{equation}
\begin{array}{ll}
\underset{\mathbf{W}}{\mathsf{minimize}} & \text{Tr}\left(\mathbf{H}\mathbf{W}\right)\\
\mathsf{subject\:to} & \text{Tr}\left(\mathbf{M}_{0}\mathbf{W}\right)=\nu\\
 & \text{Tr}\left(\mathbf{E}\mathbf{W}\right)=1\\
 & \mathbf{W}\succeq\mathbf{0}.
\end{array}\label{eq:pre and cro statistics with eq1-SDP}
\end{equation}

Like before, the nonconvex problem in \eqref{eq:pre and cro statistics with eq1}
has no duality gap. Besides the above SDP method, here we introduce
an efficient solving approach by reformulating \eqref{eq:pre and cro statistics with eq1}
into a generalized trust region subproblem (GTRS) \cite{More1993}.
Considering $\mathbf{w}=\mathbf{w}_{0}+\mathbf{F}\mathbf{x}$ where
$\mathbf{w}_{0}$ is any vector satisfying $\mathbf{1}^{T}\mathbf{w}_{0}=1$
and $\mathbf{F}$ is the kernel of $\mathbf{1}^{T}$ satisfying $\mathbf{1}^{T}\mathbf{F}=\mathbf{0}$
and a semi-unitary matrix satisfying $\mathbf{F}^{T}\mathbf{F}=\mathbf{I}$.
Let us define $\mathbf{N}=\mathbf{F}^{T}\mathbf{H}\mathbf{F}$, $\mathbf{p}=\mathbf{F}^{T}\mathbf{H}\mathbf{w}_{0}$,
$b=\mathbf{w}_{0}^{T}\mathbf{H}\mathbf{w}_{0}$, $\mathbf{N}_{0}=\mathbf{F}^{T}\mathbf{M}_{0}\mathbf{F}$
which is positive definite, $\mathbf{p}_{0}=\mathbf{F}^{T}\mathbf{M}_{0}\mathbf{w}_{0}$,
and $b_{0}=\mathbf{w}_{0}^{T}\mathbf{M}_{0}\mathbf{w}_{0}$, then
the problem in \eqref{eq:pre and cro statistics with eq1} is equivalent
to the following problem:

\begin{equation}
\begin{array}{ll}
\underset{\mathbf{x}}{\mathsf{minimize}} & \mathbf{x}^{T}\mathbf{N}\mathbf{x}+2\mathbf{p}^{T}\mathbf{x}+b\\
\mathsf{subject\:to} & \mathbf{x}^{T}\mathbf{N}_{0}\mathbf{x}+2\mathbf{p}_{0}^{T}\mathbf{x}+b_{0}=\nu,
\end{array}\label{eq:GTRS}
\end{equation}
which is a nonconvex QCQP and QCQPs of this type are specially named
GTRSs. Such problems are usually nonconvex but possess necessary and
sufficient optimality conditions based on which efficient solving
methods can be derived. We first introduce the following useful theorem.%

\begin{thm}[{\cite[Theorem 3.2]{More1993}}]
\label{thm:QCQP-global-minimizer} Consider the following QCQP:
\begin{equation}
\begin{array}{ll}
\underset{\mathbf{x}}{\mathsf{minimize}} & q\left(\mathbf{x}\right)\triangleq\mathbf{x}^{T}\mathbf{A}\mathbf{x}+2\mathbf{a}^{T}\mathbf{x}+a\\
\mathsf{subject\:to} & c\left(\mathbf{x}\right)\triangleq\mathbf{x}^{T}\mathbf{B}\mathbf{x}+2\mathbf{b}^{T}\mathbf{x}+b=0.
\end{array}\label{eq:QCQP problem}
\end{equation}
Assume that the constraint set $c\left(\mathbf{x}\right)$ is nonempty
and that $\nabla^{2}c\left(\mathbf{x}\right)=2\mathbf{B}\neq\mathbf{0}$.
A vector $\mathbf{x}^{\star}$ is a global minimizer of the problem
\eqref{eq:QCQP problem} together with a multiplier $\xi^{\star}$
if and only if the following conditions are satisfied: 
\[
\begin{cases}
\nabla q\left(\mathbf{x}^{\star}\right)+\xi^{\star}\nabla c\left(\mathbf{x}^{\star}\right)=\mathbf{0}\\
c\left(\mathbf{x}^{\star}\right)=0\\
\nabla^{2}q\left(\mathbf{x}^{\star}\right)+\xi^{\star}\nabla^{2}c\left(\mathbf{x}^{\star}\right)\succeq\mathbf{0},
\end{cases}
\]
and the interval set defined by 
\[
{\cal I}=\left\{ \xi\mid\mathbf{A}+\xi\mathbf{B}\succ\mathbf{0}\right\} 
\]
is not empty. 
\end{thm}
According to Theorem \ref{thm:QCQP-global-minimizer}, the optimality
conditions for the primal and dual variables $\left(\mathbf{x}^{\star},\xi^{\star}\right)$
of problem \eqref{eq:GTRS} are given as follows:%

\begin{equation}
\begin{cases}
\left(\mathbf{N}+\xi^{\star}\mathbf{N}_{0}\right)\mathbf{x}^{\star}+\mathbf{p}+\xi^{\star}\mathbf{p}_{0}=0\\
\mathbf{x}^{\star T}\mathbf{N}_{0}\mathbf{x}^{\star}+2\mathbf{p}_{0}^{T}\mathbf{x}^{\star}+b_{0}-\nu=0\\
\mathbf{N}+\xi^{\star}\mathbf{N}_{0}\succeq\mathbf{0}.
\end{cases}
\end{equation}
We assume $\mathbf{N}+\xi\mathbf{N}_{0}\succ\mathbf{0}$\footnote{The limiting case $\mathbf{N}+\xi\mathbf{N}_{0}$ being singular (i.e.,
$\xi=-\lambda_{\min}\left(\mathbf{N},\mathbf{N}_{0}\right)$) can
be treated separately. The assumption here is reasonable since the
case when $\xi=-\lambda_{\min}\left(\mathbf{N},\mathbf{N}_{0}\right)$
is very rare to occur theoretically and practically. }, then we can see that the optimal solution is given by
\begin{equation}
\mathbf{x}\left(\xi\right)=-\left(\mathbf{N}+\xi\mathbf{N}_{0}\right)^{-1}\left(\mathbf{p}+\xi\mathbf{p}_{0}\right),\label{eq:x_xi}
\end{equation}
and $\xi$ is the unique solution of the following equation with definition
on the interval ${\cal I}$:
\begin{equation}
\phi\left(\xi\right)=0,\:\xi\in{\cal I},
\end{equation}
where the function $\phi\left(\xi\right)$ is defined by 
\begin{equation}
\phi\left(\xi\right)=\mathbf{x}\left(\xi\right)^{T}\mathbf{N}_{0}\mathbf{x}\left(\xi\right)+2\mathbf{p}_{0}^{T}\mathbf{x}\left(\xi\right)+b_{0}-\nu,\label{eq:phi_xi}
\end{equation}
and the interval ${\cal I}$ consists of all $\xi$ for which $\mathbf{N}+\xi\mathbf{N}_{0}\succ\mathbf{0}$,
which implies that
\begin{equation}
{\cal I}=\left(-\lambda_{\min}\left(\mathbf{N},\mathbf{N}_{0}\right),\infty\right),
\end{equation}
where $\lambda_{\min}\left(\mathbf{N},\mathbf{N}_{0}\right)$ is the
minimum generalized eigenvalue of matrix pair $\left(\mathbf{N},\mathbf{N}_{0}\right)$.
\begin{thm}[{\cite[Theorem 5.2]{More1993}}]
\label{thm:phi is strictly decreasing} Assume ${\cal I}$ is not
empty, then the function $\phi\left(\xi\right)$ is strictly decreasing
on ${\cal I}$ unless $\mathbf{x}\left(\xi\right)$ is constant on
${\cal I}$.
\end{thm}
In practice, the case $\mathbf{x}\left(\xi\right)$ is constant on
${\cal I}$ cannot happen. So from Theorem \eqref{thm:phi is strictly decreasing},
we know when $\phi\left(\xi\right)$ is strictly decreasing on ${\cal I}$,
then a simple line search algorithm like bisection algorithm can be
used to find the optimal $\xi$ over ${\cal I}$. The algorithm for
problem \eqref{eq:pre and cro statistics with eq1} is summarized
in Algorithm \ref{alg:GTRS---Algorithm}.%

\begin{algorithm}[tbh]
\begin{algorithmic}[1] 

\REQUIRE $\mathbf{N}$, $\mathbf{N}_{0}$, $\mathbf{p}$, $\mathbf{p}_{0}$, $b_{0}$, $\lambda_{min}\left(\mathbf{N},\mathbf{N}_{0}\right)$, and $\nu>0$.

\STATE Set $k=0$, and choose $\xi^{\left(k\right)}\in\left(-\left(\mathbf{N},\mathbf{N}_{0}\right),\infty\right)$;

\REPEAT
\STATE Compute $\phi\left(\xi^{\left(k\right)}\right)$ according to \eqref{eq:phi_xi};
\STATE Update $\xi^{\left(k+1\right)}$ according to the value of $\phi\left(\xi^{\left(k\right)}\right)$ by a line search algorithm;
\STATE $k=k+1$;
\UNTIL convergence 
\STATE Compute $\mathbf{x}$ according to \eqref{eq:x_xi}. 

\end{algorithmic} 

\protect\caption{\label{alg:GTRS---Algorithm}GTRS - Algorithm for MRP design problems
using $\mathrm{pre}_{z}\left(\mathbf{w}\right)$ and $\mathrm{cro}_{z}\left(\mathbf{w}\right)$
with $\mathbf{w}\in\mathcal{W}_{1}$.}
\end{algorithm}

\section{Problem Solving Algorithms via Majorization-Minimization Method\label{sec:Problem-Solving-Methods-MM}}

In this section, we first discuss the majorization-minimization or
minorization-maximization (MM) method briefly, and then solving algorithms
for the MRP design problem formulations using $\mathrm{por}_{z}\left(p,\mathbf{w}\right)$
(i.e., \eqref{eq:MRP design problem formulation} with $\xi=0$ and
$\zeta=\eta=1$) and $\mathrm{pcro}_{z}\left(p,\mathbf{w}\right)$
(i.e., \eqref{eq:MRP design problem formulation} with $\xi=1$, $\mathbf{H}=\mathbf{M}_{1}$,
$\zeta=0$ and $\eta>0$) are derived based on the MM framework and
the GEVP and GTRS algorithms mentioned in the previous section.

\subsection{The MM Method\label{sub:The-MM-Method}}

The MM method \cite{HunterLange2004,RazaviyaynHongLuo2013,SunBabuPalomar2016}
refers to the majorization-minimization or minorization-maximization
which is a generalization of the well-known expectation-maximization
(EM) algorithm. The idea behind MM is that instead of dealing with
the original optimization problem which could be difficult to tackle
directly, it solves a series of simple surrogate subproblems.

Suppose the optimization problem is 
\begin{equation}
\begin{array}{ll}
\underset{\mathbf{x}}{\mathsf{minimize}} & f\left(\mathbf{x}\right)\\
\mathsf{subject\:to} & \mathbf{x}\in{\cal X},
\end{array}\label{eq:MM}
\end{equation}
where the constraint set ${\cal X}\subseteq\mathbb{R}^{N}$. In general,
there is no assumption about the convexity and differentiability on
$f\left(\mathbf{x}\right)$. The MM method aims to solve this problem
by optimizing a sequence of surrogate functions that majorize the
objective function $f\left(\mathbf{x}\right)$ over the set ${\cal X}$.
More specifically, starting from an initial feasible point $\mathbf{x}^{\left(0\right)}$,
the algorithm produces a sequence $\left\{ \mathbf{x}^{\left(k\right)}\right\} $
according to the following update rule:
\begin{equation}
\mathbf{x}^{\left(k+1\right)}\in\arg\underset{\mathbf{x}\in{\cal X}}{\min}\:u\left(\mathbf{x},\mathbf{x}^{\left(k\right)}\right),\label{eq:MM update}
\end{equation}
where $\mathbf{x}^{\left(k\right)}$ is the point generated by the
update rule at the $k$th iteration and the surrogate function $u\left(\mathbf{x},\mathbf{x}^{\left(k\right)}\right)$
is the corresponding majorizing function of $f\left(\mathbf{x}\right)$
at point $\mathbf{x}^{\left(k\right)}$. A surrogate function is called
a majorizing function of $f\left(\mathbf{x}\right)$ at point $\mathbf{x}^{\left(k\right)}$
if it satisfies the following properties:
\begin{equation}
\begin{array}{cc}
u\left(\mathbf{x},\mathbf{x}^{\left(k\right)}\right)\geq f\left(\mathbf{x}\right), & \forall\mathbf{x}\in{\cal X},\\
u\left(\mathbf{x}^{\left(k\right)},\mathbf{x}^{\left(k\right)}\right)=f\left(\mathbf{x}^{\left(k\right)}\right).
\end{array}\label{eq:majorization property}
\end{equation}
That is to say, the surrogate function $u\left(\mathbf{x},\mathbf{x}^{\left(k\right)}\right)$
should be an upper bound of the original function $f\left(\mathbf{x}\right)$
over ${\cal X}$ and coincide with $f\left(\mathbf{x}\right)$ at
point $\mathbf{x}^{\left(k\right)}$. Although the definition of $u\left(\mathbf{x},\mathbf{x}^{\left(k\right)}\right)$
gives us a great deal of flexibility for choosing it, in practice,
the surrogate function $u\left(\mathbf{x},\mathbf{x}^{\left(k\right)}\right)$
must be properly chosen so as to make the iterative update in \eqref{eq:MM update}
easy to compute while maintaining a fast convergence over the iterations.

The MM method iteratively runs until some convergence criterion is
met. Under this MM method, the objective function value is decreased
monotonically in every iteration, i.e.,
\begin{equation}
f\left(\mathbf{x}^{\left(k+1\right)}\right)\leq u\left(\mathbf{x}^{\left(k+1\right)},\mathbf{x}^{\left(k\right)}\right)\leq u\left(\mathbf{x}^{\left(k\right)},\mathbf{x}^{\left(k\right)}\right)=f\left(\mathbf{x}^{\left(k\right)}\right).\label{eq:MM convergence}
\end{equation}
The first inequality and the third equality follow from the first
and second properties of the majorizing function in \eqref{eq:majorization property}
respectively and the second inequality follows from \eqref{eq:MM update}.

\subsection{IRGEVP and IRGTRS - Solving Algorithms for MRP Design Using $\mathrm{por}_{z}\left(p,\mathbf{w}\right)$
and $\mathrm{pcro}_{z}\left(p,\mathbf{w}\right)$ }

We rewrite the problems using $\mathrm{por}_{z}\left(p,\mathbf{w}\right)$
and $\mathrm{pcro}_{z}\left(p,\mathbf{w}\right)$ in the general formulation
as follows:

\begin{equation}
\begin{array}{ll}
\underset{\mathbf{w}}{\mathsf{minimize}} & \xi\mathbf{w}^{T}\mathbf{M}_{1}\mathbf{w}+\zeta\left(\mathbf{w}^{T}\mathbf{M}_{1}\mathbf{w}\right)^{2}\\
 & +\eta\sum_{i=2}^{p}\left(\mathbf{w}^{T}\mathbf{M}_{i}\mathbf{w}\right)^{2}\\
\mathsf{subject\:to} & \mathbf{w}^{T}\mathbf{M}_{0}\mathbf{w}=\nu\\
 & \mathbf{w}\in{\cal W},
\end{array}\label{eq:por and pcro with eq01}
\end{equation}
where the specific portfolio weight constraints are implicitly replaced
by ${\cal W}$. 

To solve the problems in \eqref{eq:por and pcro with eq01} via majorization-minimization,
the key step is to find a majorizing function of the objective function
such that the majorized subproblem is easy to solve. Observe that
the objective function is quartic in $\mathbf{w}$. The following
mathematical manipulations are necessary. We first compute the Cholesky
decomposition of $\mathbf{M}_{0}$ which is $\mathbf{M}_{0}=\mathbf{L}\mathbf{L}^{T}$,
where $\mathbf{L}$ is a lower triangular with positive diagonal elements.
Let us define $\bar{\mathbf{w}}=\mathbf{L}^{T}\mathbf{w}$, $\bar{\mathbf{M}}_{i}=\mathbf{L}^{-1}\mathbf{M}_{i}\mathbf{L}^{-T}$,
and $\bar{\mathbf{W}}=\bar{\mathbf{w}}\bar{\mathbf{w}}^{T}$. The
portfolio weight set ${\cal W}$ is mapped to $\bar{{\cal W}}$ under
the linear transformation $\mathbf{L}$. Then problem \eqref{eq:por and pcro with eq01}
can be written as 

\begin{equation}
\begin{array}{ll}
\underset{\bar{\mathbf{w}},\bar{\mathbf{W}}}{\mathsf{minimize}} & \xi\text{Tr}\left(\bar{\mathbf{M}}_{1}\bar{\mathbf{W}}\right)+\zeta\left(\text{Tr}\left(\bar{\mathbf{M}}_{1}\bar{\mathbf{W}}\right)\right)^{2}\\
 & +\eta\sum_{i=1}^{p}\left(\text{Tr}\left(\bar{\mathbf{M}}_{i}\bar{\mathbf{W}}\right)\right)^{2}\\
\mathsf{subject\:to} & \bar{\mathbf{W}}=\bar{\mathbf{w}}\bar{\mathbf{w}}^{T}\\
 & \bar{\mathbf{w}}^{T}\bar{\mathbf{w}}=\nu\\
 & \bar{\mathbf{w}}\in\bar{{\cal W}}.
\end{array}\label{eq:por and pcro with eq01-SDP}
\end{equation}
Since $\text{Tr}\left(\bar{\mathbf{M}}_{i}\bar{\mathbf{W}}\right)=\text{vec}\left(\bar{\mathbf{M}}_{i}\right)^{T}\text{vec}\left(\bar{\mathbf{W}}\right)$
(recall the $\mathbf{M}_{i}$s are assumed symmetric), problem \eqref{eq:por and pcro with eq01}
can be reformulated as

\begin{equation}
\begin{array}{ll}
\underset{\bar{\mathbf{w}},\bar{\mathbf{W}}}{\mathsf{minimize}} & \xi\text{vec}\left(\bar{\mathbf{M}}_{1}\right)^{T}\text{vec}\left(\bar{\mathbf{W}}\right)+\text{vec}\left(\bar{\mathbf{W}}\right)^{T}\bar{\mathbf{M}}\text{vec}\left(\bar{\mathbf{W}}\right)\\
\mathsf{subject\:to} & \bar{\mathbf{W}}=\bar{\mathbf{w}}\bar{\mathbf{w}}^{T}\\
 & \bar{\mathbf{w}}^{T}\bar{\mathbf{w}}=\nu\\
 & \bar{\mathbf{w}}\in\bar{{\cal W}},
\end{array}\label{eq:por and pcro with eq01-SDP 2}
\end{equation}
where in the objective function 
\begin{equation}
\begin{array}{c}
\bar{\mathbf{M}}=\zeta\text{vec}\left(\bar{\mathbf{M}}_{1}\right)\text{vec}\left(\bar{\mathbf{M}}_{1}\right)^{T}+\eta\sum_{i=2}^{p}\text{vec}\left(\bar{\mathbf{M}}_{i}\right)\text{vec}\left(\bar{\mathbf{M}}_{i}\right)^{T}.\end{array}\label{eq:M_bar}
\end{equation}
Specifically, we can have the expressions for portmanteau statistics
$\mathrm{por}_{z}\left(p,\mathbf{w}\right)$ (i.e., $\zeta=1$ and
$\eta=1$) and penalized crossing statistics $\mathrm{pcro}_{z}\left(p,\mathbf{w}\right)$
(i.e., $\zeta=0$ and $\eta>0$) as follows:
\begin{equation}
\begin{cases}
\begin{array}{ccl}
\bar{\mathbf{M}}_{\mathrm{por}_{z}} & = & \sum_{i=1}^{p}\text{vec}\left(\bar{\mathbf{M}}_{i}\right)\text{vec}\left(\bar{\mathbf{M}}_{i}\right)^{T}\\
 & = & \left(\mathbf{L}\otimes\mathbf{L}\right)^{-1}\sum_{i=1}^{p}\text{vec}\left(\mathbf{M}_{i}\right)\cdot\\
 &  & \text{vec}\left(\mathbf{M}_{i}\right)^{T}\left(\mathbf{L}\otimes\mathbf{L}\right)^{-T}
\end{array}\\
\begin{array}{ccl}
\bar{\mathbf{M}}_{\mathrm{pcro}_{z}} & = & \eta\sum_{i=2}^{p}\text{vec}\left(\bar{\mathbf{M}}_{i}\right)\text{vec}\left(\bar{\mathbf{M}}_{i}\right)^{T}\\
 & = & \eta\left(\mathbf{L}\otimes\mathbf{L}\right)^{-1}\sum_{i=2}^{p}\text{vec}\left(\mathbf{M}_{i}\right)\cdot\\
 &  & \text{vec}\left(\mathbf{M}_{i}\right)^{T}\left(\mathbf{L}\otimes\mathbf{L}\right)^{-T}.
\end{array}
\end{cases}
\end{equation}

Now, the objective function in \eqref{eq:por and pcro with eq01-SDP 2}
is a quadratic function of $\bar{\mathbf{W}}$, however, this problem
is still hard to solve due to the rank-1 constraint $\bar{\mathbf{W}}=\bar{\mathbf{w}}\bar{\mathbf{w}}^{T}$.
We then consider the application of the MM trick on this problem \eqref{eq:por and pcro with eq01-SDP 2}
based on the following simple result.
\begin{lem}[{\cite[Lemma 1]{SongBabuPalomar2015}}]
\label{lem:Lemma majorization for por and pcro} Let $\mathbf{A}\in\mathbb{S}^{K}$
and $\mathbf{B}\in\mathbb{S}^{K}$ such that $\mathbf{B}\succeq\mathbf{A}$.
Then for any point $\mathbf{x}_{0}\in\mathbb{R}^{K}$, the quadratic
function \textbf{$\mathbf{x}^{T}\mathbf{A}\mathbf{x}$} is majorized
by $\mathbf{x}^{T}\mathbf{B}\mathbf{x}+2\mathbf{x}_{0}^{T}\left(\mathbf{A}-\mathbf{B}\right)\mathbf{x}+\mathbf{x}_{0}^{T}\left(\mathbf{B}-\mathbf{A}\right)\mathbf{x}_{0}$
at $\mathbf{x}_{0}$.
\end{lem}
According to Lemma \ref{lem:Lemma majorization for por and pcro},
given $\bar{\mathbf{W}}^{\left(k\right)}$ at the $k$th iteration,
we know the second part in the objective function of problem \eqref{eq:por and pcro with eq01-SDP 2}
is majorized by the following majorizing function at $\bar{\mathbf{W}}^{\left(k\right)}$:

\begin{equation}
\begin{array}{rl}
 & u_{1}\left(\bar{\mathbf{W}},\bar{\mathbf{W}}^{\left(k\right)}\right)\\
= & \psi\left(\bar{\mathbf{M}}\right)\text{vec}\left(\bar{\mathbf{W}}\right)^{T}\text{vec}\left(\bar{\mathbf{W}}\right)\\
 & +2\text{vec}\left(\bar{\mathbf{W}}^{\left(k\right)}\right)^{T}\left(\bar{\mathbf{M}}-\psi\left(\bar{\mathbf{M}}\right)\mathbf{I}\right)\text{vec}\left(\bar{\mathbf{W}}\right)\\
 & +\text{vec}\left(\bar{\mathbf{W}}^{\left(k\right)}\right)^{T}\left(\psi\left(\bar{\mathbf{M}}\right)\mathbf{I}-\bar{\mathbf{M}}\right)\text{vec}\left(\bar{\mathbf{W}}^{\left(k\right)}\right),
\end{array}\label{eq:majorization function for por and pcro}
\end{equation}
where $\psi\left(\bar{\mathbf{M}}\right)$ is a scalar number depending
on $\bar{\mathbf{M}}$ and satisfying $\psi\left(\bar{\mathbf{M}}\right)\mathbf{I}\succeq\bar{\mathbf{M}}$.
Since the first term $\text{vec}\left(\bar{\mathbf{W}}\right)^{T}\text{vec}\left(\bar{\mathbf{W}}\right)=\left(\bar{\mathbf{w}}^{T}\bar{\mathbf{w}}\right)^{2}=\nu^{2}$
and the last term only depends on $\bar{\mathbf{W}}^{\left(k\right)}$,
they are just two constants. 

On the choice of $\psi\left(\bar{\mathbf{M}}\right)$, according to
Lemma \ref{lem:Lemma majorization for por and pcro}, it is obvious
to see that $\psi\left(\bar{\mathbf{M}}\right)$ can be easily chosen
to be $\lambda_{\max}\left(\bar{\mathbf{M}}\right)=\left\Vert \bar{\mathbf{M}}\right\Vert _{2}$.
In the implementation of the algorithm, although $\left\Vert \bar{\mathbf{M}}\right\Vert _{2}$
only needs to be computed once for the whole algorithm, it is still
not computationally easy to get. In view of this, we introduce the
following lemma to obtain more possibilities for $\psi\left(\bar{\mathbf{M}}\right)$
which could be relatively easy to compute.
\begin{lem}[\cite{HornJohnson2012}]
\label{lem:majorization scaler for M} For any matrix $\mathbf{B}\in\mathbb{R}^{P\times Q}$,
the following inequalities about $\left\Vert \mathbf{B}\right\Vert _{2}$
hold:%

\[
\begin{aligned} & \left\Vert \mathbf{B}\right\Vert _{2}\\
 & \leq\begin{cases}
\left\Vert \mathbf{B}\right\Vert _{F}=\sqrt{\sum_{i=1}^{P}\sum_{j=1}^{Q}\left|b_{ij}\right|^{2}}\\
\sqrt{P}\left\Vert \mathbf{B}\right\Vert _{\infty}=\sqrt{P}\max_{i=1,\ldots,P}\sum_{j=1}^{Q}\left|b_{ij}\right|\\
\sqrt{Q}\left\Vert \mathbf{B}\right\Vert _{1}=\sqrt{Q}\max_{j=1,\ldots,Q}\sum_{i=1}^{P}\left|b_{ij}\right|\\
\sqrt{PQ}\left\Vert \mathbf{B}\right\Vert _{max}=\sqrt{PQ}\max_{i=1,\ldots,P}\max_{j=1,\ldots,Q}\left|b_{ij}\right|\\
\sqrt{\left\Vert \mathbf{B}\right\Vert _{\infty}\left\Vert \mathbf{B}\right\Vert _{1}}\\
=\sqrt{\left(\max_{i=1,\ldots,P}\sum_{j=1}^{Q}\left|b_{ij}\right|\right)\left(\max_{j=1,\ldots,Q}\sum_{i=1}^{P}\left|b_{ij}\right|\right)}.
\end{cases}
\end{aligned}
\]

\end{lem}
According to the above relations, $\psi\left(\bar{\mathbf{M}}\right)$
can be chosen to be any number is larger than $\left\Vert \bar{\mathbf{M}}\right\Vert _{2}$
but much easier to compute.

After ignoring the constants in \eqref{eq:majorization function for por and pcro},
the majorized problem of problem \eqref{eq:por and pcro with eq01-SDP 2}
is given by 

\begin{equation}
\begin{array}{ll}
\underset{\bar{\mathbf{w}},\bar{\mathbf{W}}}{\mathsf{minimize}} & \xi\text{vec}\left(\bar{\mathbf{M}}_{1}\right)^{T}\text{vec}\left(\bar{\mathbf{W}}\right)\\
 & +2\text{vec}\left(\bar{\mathbf{W}}^{\left(k\right)}\right)^{T}\left(\bar{\mathbf{M}}-\psi\left(\bar{\mathbf{M}}\right)\mathbf{I}\right)\text{vec}\left(\bar{\mathbf{W}}\right)\\
\mathsf{subject\:to} & \bar{\mathbf{W}}=\bar{\mathbf{w}}\bar{\mathbf{w}}^{T}\\
 & \bar{\mathbf{w}}^{T}\bar{\mathbf{w}}=\nu\\
 & \bar{\mathbf{w}}\in\bar{{\cal W}},
\end{array}\label{eq:por and pcro statistics with eq01-MM}
\end{equation}
which can be further written as 

\begin{equation}
\begin{array}{ll}
\underset{\bar{\mathbf{w}},\bar{\mathbf{W}}}{\mathsf{minimize}} & \xi\text{Tr}\left(\bar{\mathbf{M}}_{1}\bar{\mathbf{W}}\right)+2\zeta\text{Tr}\left(\bar{\mathbf{M}}_{1}\bar{\mathbf{W}}^{\left(k\right)}\right)\text{Tr}\left(\bar{\mathbf{M}}_{1}\bar{\mathbf{W}}\right)\\
 & +2\eta\sum_{i=2}^{p}\text{Tr}\left(\bar{\mathbf{M}}_{i}\bar{\mathbf{W}}^{\left(k\right)}\right)\text{Tr}\left(\bar{\mathbf{M}}_{i}\bar{\mathbf{W}}\right)\\
 & -2\psi\left(\bar{\mathbf{M}}\right)\text{Tr}\left(\bar{\mathbf{W}}^{\left(k\right)}\bar{\mathbf{W}}\right)\\
\mathsf{subject\:to} & \bar{\mathbf{W}}=\bar{\mathbf{w}}\bar{\mathbf{w}}^{T}\\
 & \bar{\mathbf{w}}^{T}\bar{\mathbf{w}}=\nu\\
 & \bar{\mathbf{w}}\in\bar{{\cal W}}.
\end{array}\label{eq:por and pcro statistics with eq01-MM 2}
\end{equation}

By changing $\bar{\mathbf{W}}$ back to $\bar{\mathbf{w}}$, problem
\eqref{eq:por and pcro statistics with eq01-MM 2} becomes%
{} 
\begin{equation}
\begin{array}{ll}
\underset{\bar{\mathbf{w}}}{\mathsf{minimize}} & \bar{\mathbf{w}}^{T}\bar{\mathbf{H}}^{\left(k\right)}\bar{\mathbf{w}}\\
\mathsf{subject\:to} & \bar{\mathbf{w}}^{T}\bar{\mathbf{w}}=\nu\\
 & \bar{\mathbf{w}}\in\bar{{\cal W}},
\end{array}\label{eq:por and pcro with eq01-subproblem in w_bar}
\end{equation}
where in the objective function, $\bar{\mathbf{H}}^{\left(k\right)}$
is defined in this way $\bar{\mathbf{H}}^{\left(k\right)}=\xi\bar{\mathbf{M}}_{1}+2\zeta\left(\bar{\mathbf{w}}^{\left(k\right)T}\bar{\mathbf{M}}_{1}\bar{\mathbf{w}}^{\left(k\right)}\right)\bar{\mathbf{M}}_{1}+2\eta\sum_{i=2}^{p}\left(\bar{\mathbf{w}}^{\left(k\right)T}\bar{\mathbf{M}}_{i}\bar{\mathbf{w}}^{\left(k\right)}\right)\bar{\mathbf{M}}_{i}-2\psi\left(\bar{\mathbf{M}}\right)\bar{\mathbf{w}}^{\left(k\right)}\bar{\mathbf{w}}^{\left(k\right)T}$.
Finally, we can undo the change of variable $\bar{\mathbf{w}}=\mathbf{L}^{T}\mathbf{w}$,
obtaining 

\begin{equation}
\begin{array}{ll}
\underset{\mathbf{w}}{\mathsf{minimize}} & \mathbf{w}^{T}\mathbf{H}^{\left(k\right)}\mathbf{w}\\
\mathsf{subject\:to} & \mathbf{w}^{T}\mathbf{M}_{0}\mathbf{w}=\nu\\
 & \mathbf{w}\in{\cal W},
\end{array}\label{eq:por and pcro with eq01-subproblem in w}
\end{equation}
where in the objective function 
\begin{equation}
\begin{array}{ccl}
\mathbf{H}^{\left(k\right)} & = & \xi\mathbf{M}_{1}+2\zeta\left(\mathbf{w}^{\left(k\right)T}\mathbf{M}_{1}\mathbf{w}^{\left(k\right)}\right)\mathbf{M}_{1}\\
 &  & +2\eta\sum_{i=2}^{p}\left(\mathbf{w}^{\left(k\right)T}\mathbf{M}_{i}\mathbf{w}^{\left(k\right)}\right)\mathbf{M}_{i}\\
 &  & -2\psi\left(\bar{\mathbf{M}}\right)\mathbf{M}_{0}\mathbf{w}^{\left(k\right)}\mathbf{w}^{\left(k\right)T}\mathbf{M}_{0}.
\end{array}\label{eq:H in MM}
\end{equation}
 More specifically, for portmanteau statistics $\mathrm{por}_{z}\left(p,\mathbf{w}\right)$
(i.e., $\xi=0$, $\zeta=1$ and $\eta=1$) and penalized crossing
statistics $\mathrm{pcro}_{z}\left(p,\mathbf{w}\right)$ (i.e., $\xi=1$,
$\zeta=0$ and $\eta>0$), we have the following expressions:
\begin{equation}
\begin{cases}
\begin{array}{rcl}
\mathbf{H}_{\mathrm{por}_{z}}^{\left(k\right)} & = & 2\sum_{i=1}^{p}\left(\mathbf{w}^{\left(k\right)T}\mathbf{M}_{i}\mathbf{w}^{\left(k\right)}\right)\mathbf{M}_{i}\\
 &  & -2\psi\left(\bar{\mathbf{M}}\right)\mathbf{M}_{0}\mathbf{w}^{\left(k\right)}\mathbf{w}^{\left(k\right)T}\mathbf{M}_{0},
\end{array}\\
\begin{array}{rcl}
\mathbf{H}_{\mathrm{pcro}_{z}}^{\left(k\right)} & = & \mathbf{M}_{1}+2\eta\sum_{i=2}^{p}\left(\mathbf{w}^{\left(k\right)T}\mathbf{M}_{i}\mathbf{w}^{\left(k\right)}\right)\mathbf{M}_{i}\\
 &  & -2\psi\left(\bar{\mathbf{M}}\right)\mathbf{M}_{0}\mathbf{w}^{\left(k\right)}\mathbf{w}^{\left(k\right)T}\mathbf{M}_{0}.
\end{array}
\end{cases}
\end{equation}

Finally, in the majorization problems \eqref{eq:por and pcro with eq01-subproblem in w_bar}
and \eqref{eq:por and pcro with eq01-subproblem in w}, the objective
functions become quadratic in the variable rather than quartic in
the variable as in the original problem \eqref{eq:por and pcro with eq01}.
Depending on the specific form of ${\cal W}$, problem \eqref{eq:por and pcro with eq01-subproblem in w}
is either the GEVP or GTRS problems discussed in the previous sections.
So, in order to handle the original problem \eqref{eq:por and pcro with eq01}
directly which could be difficult, we just need to iteratively solve
a sequence of GEVPs or GTRSs. We call these MM-based algorithms iteratively
reweighted GEVP (IRGEVP) and iteratively reweighted GTRS (IRGTRS)
respectively which are summarized in Algorithm \ref{alg:IRGEVP-and-IRGTRS algorithm}.%

\begin{algorithm}[tbh]
\begin{algorithmic}[1] 
\REQUIRE $p$,  $\mathbf{M}_{i}$ with $i=1,\ldots,p$, and $\nu>0$. 
\STATE Set $k=0$, and choose initial value $\mathbf{w}^{\left(k\right)}\in{\cal W}$;
\STATE Compute $\bar{\mathbf{M}}$ according to \eqref{eq:M_bar} and $ \psi\left(\bar{\mathbf{M}}\right)$;

\REPEAT
\STATE Compute $\mathbf{H}^{\left(k\right)}$ according to \eqref{eq:H in MM};
\STATE Update $\mathbf{w}^{\left(k+1\right)}$ by solving the GEVP in \eqref{eq:GEVP} or the GTRS in \eqref{eq:GTRS};
\STATE $k=k+1$;
\UNTIL convergence 
\end{algorithmic} \protect\caption{\label{alg:IRGEVP-and-IRGTRS algorithm}IRGEVP and IRGTRS - Algorithms
for MRP design problems using $\mathrm{por}_{z}\left(p,\mathbf{w}\right)$
and $\mathrm{pcro}_{z}\left(p,\mathbf{w}\right)$.}
\end{algorithm}

\subsection{EIRGEVP - An Extended Algorithm for IRGEVP}

In the MM-based algorithms mentioned above, it would be much desirable
if we could get a closed-form solution for the subproblems in every
iteration. In fact, for IRGEVPs, applying the MM trick once again,
a closed-form solution is attainable at every iteration. To illustrate
this, we rewrite the subproblem \eqref{eq:por and pcro with eq01-subproblem in w_bar}
of IRGEVP again as follows:%

\begin{equation}
\begin{array}{ll}
\underset{\mathbf{w}}{\mathsf{minimize}} & \mathbf{w}^{T}\mathbf{H}^{\left(k\right)}\mathbf{w}\\
\mathsf{subject\:to} & \mathbf{w}^{T}\mathbf{M}_{0}\mathbf{w}=\nu\\
 & \mathbf{1}^{T}\mathbf{w}=0.
\end{array}\label{eq:closed form solution for IRGEVP-1}
\end{equation}
Considering the trick used to eliminate the linear constraint to get
problem \eqref{eq:GEVP}, we can get the following equivalent formulation: 

\begin{equation}
\begin{array}{ll}
\underset{\mathbf{x}}{\mathsf{minimize}} & \mathbf{x}^{T}\mathbf{N}^{\left(k\right)}\mathbf{x}\\
\mathsf{subject\:to} & \mathbf{x}^{T}\mathbf{N}_{0}\mathbf{x}=\nu,
\end{array}\label{eq:closed form solution for IRGEVP-1-GEVP}
\end{equation}
where $\mathbf{F}$ and $\mathbf{N}_{0}$ are defined as before; $\mathbf{N}^{\left(k\right)}=\mathbf{F}^{T}\mathbf{H}^{\left(k\right)}\mathbf{F}$.
Considering the Cholesky decomposition $\mathbf{N}_{0}=\mathbf{R}\mathbf{R}^{T}$
with $\mathbf{R}$ to be a lower triangular with positive diagonal
elements, we can have the variable transformation $\bar{\mathbf{x}}=\mathbf{R}^{T}\mathbf{x}$.
Then the problem \eqref{eq:closed form solution for IRGEVP-1} becomes

\begin{equation}
\begin{array}{ll}
\underset{\bar{\mathbf{x}}}{\mathsf{minimize}} & \bar{\mathbf{x}}^{T}\bar{\mathbf{N}}^{\left(k\right)}\bar{\mathbf{x}}\\
\mathsf{subject\:to} & \bar{\mathbf{x}}^{T}\bar{\mathbf{x}}=\nu,
\end{array}\label{eq:closed form solution for IRGEVP-2}
\end{equation}
where $\bar{\mathbf{N}}^{\left(k\right)}=\mathbf{R}^{-1}\mathbf{N}^{\left(k\right)}\mathbf{R}^{-T}$.

Applying Lemma \ref{lem:Lemma majorization for por and pcro} again,
the objective function of problem \eqref{eq:closed form solution for IRGEVP-2}
is majorized by the following majorizing function at $\bar{\mathbf{x}}^{\left(k\right)}$:

\begin{equation}
\begin{array}{rl}
 & u_{2}\left(\bar{\mathbf{x}},\bar{\mathbf{x}}^{\left(k\right)}\right)\\
= & \psi\left(\bar{\mathbf{N}}^{\left(k\right)}\right)\bar{\mathbf{x}}^{T}\bar{\mathbf{x}}\\
 & +2\left[\left(\bar{\mathbf{N}}^{\left(k\right)}-\psi\left(\bar{\mathbf{N}}^{\left(k\right)}\right)\mathbf{I}\right)\bar{\mathbf{x}}^{\left(k\right)}\right]^{T}\bar{\mathbf{x}}\\
 & +\bar{\mathbf{x}}^{\left(k\right)T}\left[\psi\left(\bar{\mathbf{N}}^{\left(k\right)}\right)\mathbf{I}-\bar{\mathbf{N}}^{\left(k\right)}\right]\bar{\mathbf{x}}^{\left(k\right)},
\end{array}\label{eq:majorization function for por and pcro with eq0}
\end{equation}
where $\psi\left(\bar{\mathbf{N}}^{\left(k\right)}\right)$ can be
chosen using the results from Lemma \ref{lem:majorization scaler for M}.
The first and last parts are just two constants. Note that although
in the derivation we have applied the MM scheme twice, it can be viewed
as a direct majorization for the objective of the original problem
at $\mathbf{w}^{\left(k\right)}$. The following lemma summarizes
the overall majorizing function. 
\begin{lem}
\label{lem:majorization-overall}For problem \eqref{eq:por and pcro with eq01}
with $\mathbf{w}\in{\cal W}_{0}$, the majorization in \eqref{eq:majorization function for por and pcro}
together with \eqref{eq:majorization function for por and pcro with eq0}
can be shown to be a majorization for the objective function of the
original problem at $\mathbf{w}^{\left(k\right)}$ over the constraint
set by the following function:

\begin{equation}
\begin{array}{rl}
 & u_{2}\left(\mathbf{w},\mathbf{w}^{\left(k\right)}\right)\\
= & 2\left[\left(\mathbf{H}^{\left(k\right)}-\psi\left(\mathbf{R}^{-1}\mathbf{F}^{T}\mathbf{H}^{\left(k\right)}\mathbf{F}\mathbf{R}^{-T}\right)\mathbf{M}_{0}\right)\mathbf{w}^{\left(k\right)}\right]^{T}\mathbf{w}\\
 & +2\psi\left(\mathbf{R}^{-1}\mathbf{F}^{T}\mathbf{H}^{\left(k\right)}\mathbf{F}\mathbf{R}^{-T}\right)\nu-\mathbf{w}^{\left(k\right)T}\mathbf{H}^{\left(k\right)}\mathbf{w}^{\left(k\right)}.
\end{array}\label{eq:majorization function (whole) for por and pcro with eq0}
\end{equation}
where the last two terms are constants.\end{lem}
\begin{IEEEproof}
See Appendix \ref{Proof-for-Lemma-majorazation-overall}.
\end{IEEEproof}
Then, the majorized problem of \eqref{eq:closed form solution for IRGEVP-2}
becomes

\begin{equation}
\begin{array}{ll}
\underset{\bar{\mathbf{x}}}{\mathsf{minimize}} & \mathbf{e}^{\left(k\right)T}\bar{\mathbf{x}}\\
\mathsf{subject\:to} & \bar{\mathbf{x}}^{T}\bar{\mathbf{x}}=\nu,
\end{array}\label{eq:closed form solution for IRGEVP-subproblem}
\end{equation}
where $\mathbf{e}^{\left(k\right)}=2\left(\bar{\mathbf{N}}^{\left(k\right)}-\psi\left(\bar{\mathbf{N}}^{\left(k\right)}\right)\mathbf{I}\right)\bar{\mathbf{x}}^{\left(k\right)}$
for the majorization in \eqref{eq:majorization function for por and pcro with eq0}.
By Cauchy-Schwartz inequality, we have $\mathbf{e}^{T}\bar{\mathbf{x}}\geq-\left\Vert \mathbf{e}\right\Vert _{2}\left\Vert \bar{\mathbf{x}}\right\Vert _{2}=-\nu\left\Vert \mathbf{e}\right\Vert _{2}$,
and the equality holds only when $\bar{\mathbf{x}}$ and $\mathbf{e}$
are aligned in the opposite direction. Considering the constraint,
we can get the optimal solution of \eqref{eq:closed form solution for IRGEVP-subproblem}
as $\bar{\mathbf{x}}=-\sqrt{\nu}\frac{\mathbf{e}}{\left\Vert \mathbf{e}\right\Vert _{2}}$.
We call this algorithm extended IRGEVP (EIRGEVP) which is summarized
in Algorithm \ref{alg:EIRGEVP_algorithm}.

\begin{algorithm}[tbh]
\begin{algorithmic}[1] 
\REQUIRE $p$,  $\mathbf{M}_{i}$ with $i=1,\ldots,p$, and $\nu>0$. 
\STATE Set $k=0$, and choose initial value $\mathbf{w}^{\left(k\right)}\in{\cal W}$;
\STATE Compute $\bar{\mathbf{M}}$ and $ \psi\left(\bar{\mathbf{M}}\right)$;

\REPEAT
\STATE Compute $\bar{\mathbf{N}}^{\left(k\right)}$ and $\psi\left(\bar{\mathbf{N}}^{\left(k\right)}\right)$;
\STATE Update $\mathbf{w}^{\left(k+1\right)}$ with a closed-form solution;
\STATE $k=k+1$;
\UNTIL convergence 
\end{algorithmic} 

\protect\caption{\label{alg:EIRGEVP_algorithm}EIRGEVP - An extended algorithm for
IRGEVP.}
\end{algorithm}

\section{Numerical Experiments\label{sec:Numerical-Experiments}}

A statistical arbitrage strategy involves several steps of which the
MRP design is a central one. Here, we divide the whole strategy into
four sequential steps, namely: assets pool construction, MRP design,
unit-root test, and mean-reversion trading. In the first step, we
select a collection of possibly cointegrated asset candidates to construct
an asset pool, on which we will not elaborate in this paper.%
{} In the second step, based on the candidate assets from the asset
pool, MRPs are designed using either traditional design methods like
Engle-Granger OLS method \cite{EngleGranger1987} and Johansen method
\cite{Johansen1988} or the proposed methods in this paper. In the
third step, unit-root test procedures like Augmented Dickey-Fuller
test \cite{DickeyFuller1979} and Phillips-Perron test \cite{PhillipsPerron1988}
are applied to test the stationarity or mean-reversion property of
the designed MRPs.%
{} In the fourth step, MRPs passing the unit-root tests will be traded
based on a designed mean-reversion trading strategy.

In this section, we first illustrate a mean-reversion trading strategy
and based on that the performance of our proposed MRP design methods
in Sections \ref{sec:Problem-Solving-Methods-GEVP-GTRS} and \ref{sec:Problem-Solving-Methods-MM}
using both synthetic data and real market data are shown accordingly.

\subsection{Mean-Reversion Trading Design}

In this paper, we use a simple trading strategy where the trading
signals, i.e., to buy, to sell, or simply to hold, are designed based
on simple event triggers.%
{} Mean-reversion trading is carried out on the designed spread $z_{t}$
which is tested to be unit-root stationary. A trading position (either
a long position denoted by $1$ or a short position denoted by $-1$)
denotes a state for investment and it is opened when the spread $z_{t}$
is away from its long-run equilibrium $\mu_{z}$ by a predefined trading
threshold $\Delta$ and closed (denoted by $0$) when $z_{t}$ crosses
its equilibrium $\mu_{z}$. (A common variation is to close the position
after the spread crossed the equilibrium by more than another threshold
$\Delta^{\prime}$.) The time period from position opening to position
closing is defined as a trading period. 

In order to get a standard trading rule, we introduce a standardization
technique by defining $\mathrm{z-score}$ which is a normalized spread
as follows:%

\begin{equation}
\tilde{z}_{t}=\frac{z_{t}-\mu_{z}}{\sigma_{z}},\label{eq:z-score}
\end{equation}
where $\mu_{z}$ and $\sigma_{z}$ are the mean and the standard deviation
of the spread $z_{t}$ and computed over an in-sample look-back period
in practice. For $\tilde{z}_{t}$, it follows that $\mathsf{E}\left[\tilde{z}_{t}\right]=0$
and $\mathsf{Std}\left[\tilde{z}_{t}\right]=1$. Then, we can define
$\Delta=d\times\sigma_{z}$, for some value of $d$ (e.g., $d=1$). 

\begin{figure}[tbh]

\begin{centering}
\includegraphics[scale=0.5]{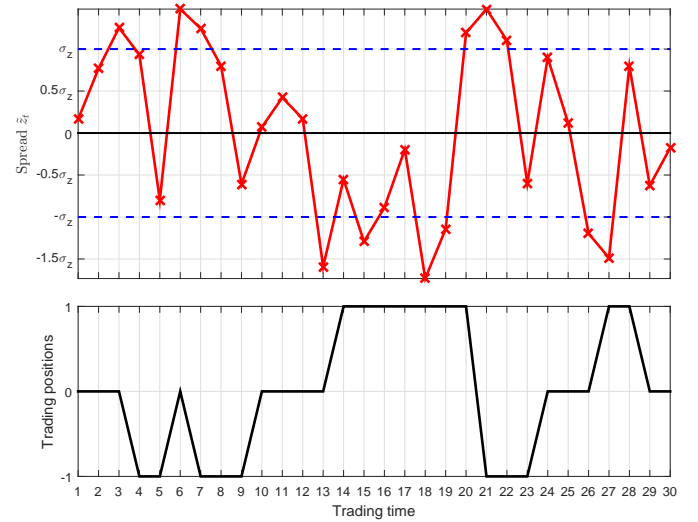}
\par\end{centering}

\begin{centering}
\protect\caption{\label{fig:mean-reversion_trading}A simple example for mean-reversion
trading strategy design (trading threshold $\Delta=\sigma_{z}$).}

\par\end{centering}

\end{figure}

In a trading stage, based on the trading position and observed (normalized)
spread value at holding period $t$, we can get the trading actions
at the next consecutive holding period $t+1$. The mean-reversion
trading strategy is summarized in Table \ref{tab:Trading-Signals-Actions}
and a simple trading example based on this strategy is illustrated
in Figure \ref{fig:mean-reversion_trading}.%

\begin{table*}[tbh]
\begin{centering}
\protect\caption{\label{tab:Trading-Signals-Actions}Trading Positions, Normalized
Spread, and Trading Actions of a Mean-Reversion Trading Strategy}

\par\end{centering}

\centering{}%
\begin{tabular}{|c||c|c|c|}
\hline 
Trading Position at $t$ & Normalized Spread $\tilde{z}_{t}$ & Action(s) Taken within Holding Period $t+1$ & Trading Position at $t+1$\tabularnewline
\hline 
\hline 
\multirow{3}{*}{1} & $+d\leq\tilde{z}_{t}$ & Close the long pos. \& Open a short pos.  & -1\tabularnewline
\cline{2-4} 
 & $0\leq\tilde{z}_{t}<+d$ & Close the long pos.  & 0\tabularnewline
\cline{2-4} 
 & $\tilde{z}_{t}<0$ & No action & 1\tabularnewline
\hline 
\multirow{3}{*}{0} & $+d\leq\tilde{z}_{t}$ & Open a short pos.  & -1\tabularnewline
\cline{2-4} 
 & $-d<\tilde{z}_{t}<+d$ & No action & 0\tabularnewline
\cline{2-4} 
 & $\tilde{z}_{t}\leq-d$ & Open a long pos.  & 1\tabularnewline
\hline 
\multirow{3}{*}{-1} & $0<\tilde{z}_{t}$ & No action & -1\tabularnewline
\cline{2-4} 
 & $-d<\tilde{z}_{t}\leq0$ & Close the short pos.  & 0\tabularnewline
\cline{2-4} 
 & $\tilde{z}_{t}\leq-d$ & Close the short pos. \& Open a long pos.  & 1\tabularnewline
\hline 
\end{tabular}
\end{table*}

\subsection{Performance Metrics}

After an MRP is constructed, we need to define the relation between
the designed MRP with the underlying financial assets. Recall that
the spread for the designed MRP is $z_{t}=\mathbf{w}^{T}\mathbf{s}_{t}$,
where $\mathbf{s}_{t}=\left[s_{1,t},s_{2,t},\ldots,s_{N,t}\right]^{T}$
with $s_{n,t}=\mathbf{w}_{s_{n}}^{T}\mathbf{y}_{t}$ for $n=1,2,\ldots,N$.
By defining $\mathbf{s}_{t}=\mathbf{W}_{s}^{T}\mathbf{y}_{t}$ with
$\mathbf{W}_{s}=\left[\mathbf{w}_{s_{1}},\mathbf{w}_{s_{2}},\dots,\mathbf{w}_{s_{N}}\right]$,
we get the spread $z_{t}=\mathbf{w}_{p}^{T}\mathbf{y}_{t}$, where
$\mathbf{w}_{p}=\mathbf{W}_{s}\mathbf{w}$ denotes the portfolio weight
directly defined on the underlying assets.

Based on the mean-reversion trading strategy introduced before and
the MRP defined by $\mathbf{w}_{p}$ here, we employ the following
performance metrics in the numerical experiments.

\subsubsection{Portfolio Return Measures}

In the following, we first give the return definition for one single
asset, and after that, several different return measures for an MRP
are talked about.

For one single asset, the return or cumulative return at time $t$
for $\tau$ holding periods is defined as $r_{t}\left(\tau\right)=\frac{p_{t}-p_{t-\tau}}{p_{t-\tau}}$,
where $\tau$ in the parentheses denotes the period length and is
usually omitted when the length is one. Here, the return $r_{t}\left(\tau\right)$
as a rate of return is used to measure the aggregate amount of profits
or losses (in percentage) of an investment strategy on one asset over
a time period $\tau$. %

\paragraph{Profit and Loss (P\&L)}

The profit and loss (P\&L) measures the amount of profits or losses
(in units of dollars) of an investment on the portfolio for some holding
periods. 

Within one trading period, if a long position is opened on an MRP
at time $t_{o}$ and closed at time $t_{c}$, then the multi-period
P\&L of this MRP at time $t$ ($t_{o}\leq t\leq t_{c}$) accumulated
from $t_{o}$ is computed as $\mathrm{P\&L}_{t}\left(\tau\right)=\mathbf{w}_{p}^{T}\mathbf{r}_{t}\left(\tau\right)=\mathbf{w}_{p}^{T}\mathbf{r}_{t}\left(t-t_{o}\right)$,
where $\tau=t-t_{o}$ denotes the length of the holding periods, and
$\mathbf{r}_{t}\left(\tau\right)=\left[r_{1,t}\left(\tau\right),r_{2,t}\left(\tau\right),\ldots,r_{M,t}\left(\tau\right)\right]^{T}$
is the return vector. More generally, the cumulative P\&L of this
MRP at time $t$ for $\tau$ ($0\leq\tau\leq t-t_{o}$) holding periods
is defined as 
\begin{equation}
\mathrm{P\&L}_{t}\left(\tau\right)=\mathbf{w}_{p}^{T}\mathbf{r}_{t}\left(t-t_{o}\right)-\mathbf{w}_{p}^{T}\mathbf{r}_{t-\tau}\left(t-\tau-t_{o}\right),\label{eq:multi-period PnL}
\end{equation}
where we define $\mathbf{r}_{t}\left(0\right)=\mathbf{0}$. Then we
have the single-period P\&L (e.g., daily P\&L, monthly P\&L) denoted
by $\mathrm{P\&L}_{t}$ at time $t$ (i.e., $\tau=1$) is computed
as
\begin{equation}
\mathrm{P\&L}_{t}=\mathbf{w}_{p}^{T}\mathbf{r}_{t}\left(t-t_{o}\right)-\mathbf{w}_{p}^{T}\mathbf{r}_{t-1}\left(t-1-t_{o}\right).\label{eq:single-period PnL}
\end{equation}

Likewise, within one trading period, if a short position is opened
on this MRP, then multi-period P\&L is $\mathrm{P\&L}_{t}\left(\tau\right)=\mathbf{w}_{p}^{T}\mathbf{r}_{t-\tau}\left(t-\tau-t_{o}\right)-\mathbf{w}_{p}^{T}\mathbf{r}_{t}\left(t-t_{o}\right)$
and the single-period P\&L is $\mathrm{P\&L}_{t}=\mathbf{w}_{p}^{T}\mathbf{r}_{t-1}\left(t-1-t_{o}\right)-\mathbf{w}_{p}^{T}\mathbf{r}_{t}\left(t-t_{o}\right)$.
About the portfolio P\&L calculation within the trading periods, we
have the following lemma. 
\begin{lem}[\textbf{P\&L Calculation for Mean-Reversion Trading}]
\label{lem:spread-return} Within one trading period, if the price
change of every asset in an MRP is small enough, then the P\&L in
\eqref{eq:multi-period PnL} can be approximately calculated by the
change of the log-price spread $z_{t}$. Specifically,

1) for a long position opened on the MRP, $\mathrm{P\&L}_{t}\left(\tau\right)\approx z_{t}-z_{t-\tau}$;
and 

2) for a short position opened on the MRP, $\mathrm{P\&L}_{t}\left(\tau\right)\approx z_{t-\tau}-z_{t}$.\end{lem}
\begin{IEEEproof}
See Appendix \ref{Proof-for-Lemma-spread-return}.
\end{IEEEproof}
This lemma reveals the philosophy of the MRP design and also the mean-reversion
trading by showing the connection between the log-price spread value
and the portfolio return. 

Since there is no trading conduct between two trading periods, the
P\&L measures (both the multi-period P\&L and single-period P\&L)
are simply defined to be 0.%

\paragraph{Cumulative P\&L}

In order to measure the cumulative return performance for an MRP,
we define the cumulative P\&L in one trading from time $t_{1}$ to
$t_{2}$ as %
{} 
\begin{equation}
\begin{array}{c}
\mathrm{Cum.\ P\&L}\left(t_{1},t_{2}\right)=\sum_{t=t_{1}}^{t_{2}}\mathrm{P\&L}_{t}.\end{array}\label{eq:cumulative PnL}
\end{equation}

\paragraph{Return on Investment (ROI)}

Since different MRPs may have different leverage properties due to
$\mathbf{w}_{p}$, we introduce another portfolio return measure (rate
of return) called return on investment (ROI). 

Within one trading period, the ROI at time $t$ ($t_{o}\leq t\leq t_{c}$)
is defined to be the single-period P\&L at time $t$ normalized by
the gross investment deployed which is $\left\Vert \mathbf{w}_{p}\right\Vert _{1}$
(that is the gross investment exposure to the market including the
long position investment and the short position investment) written
as

\begin{equation}
\mathrm{ROI}_{t}=\mathrm{P\&L}_{t}/\left\Vert \mathbf{w}_{p}\right\Vert _{1}.\label{eq:single-period ROI}
\end{equation}
Like the P\&L measures, between two trading periods, $\mathrm{ROI}_{t}$
is defined to be 0.%

\subsubsection{Sharpe Ratio (SR)}

The Sharpe ratio (SR) \cite{Sharpe1994} is a measure for calculating
risk-adjusted return. It describes how much excess return one can
receive for the extra volatility (square root of variance). 

Here, the Sharpe ratio of ROI (or, equivalently, Sharpe ratio of P\&L)
for a trading stage from time $t_{1}$ to $t_{2}$ is defined as follows:

\begin{equation}
\mathrm{SR}_{\mathrm{ROI}}\left(t_{1},t_{2}\right)=\frac{\mu_{\mathrm{ROI}}}{\sigma_{\mathrm{ROI}}},\label{eq:Sharpe ratio for ROI}
\end{equation}
where $\mu_{\mathrm{ROI}}=\frac{1}{t_{2}-t_{1}}\sum_{t=t_{1}}^{t_{2}}\mathrm{ROI}_{t}$
and $\sigma_{\mathrm{ROI}}=\left[\frac{1}{t_{2}-t_{1}}\sum_{t=t_{1}}^{t_{2}}\left(\mathrm{ROI}_{t}-\mu_{\mathrm{ROI}}\right)^{2}\right]^{1/2}$.
In the computation of the Sharpe ratio, we set the risk-free return
to $0$, in which case it reduces to the information ratio.%
{} %
{}

\subsection{Synthetic Data Experiments}

For synthetic data experiments, we generate the sample path of log-prices
for $M$ financial assets using a multivariate cointegrated systems
\cite{Tsay2005,Luetkepohl2007}, where there are $r$ long-run cointegration
relations and $M-r$ common trends. We divide the sample path into
two stages: in-sample training stage and out-of-sample backtesting
or trading stage. All the parameters like spread equilibrium $\mu_{z}$,
trading threshold $\Delta$, and portfolio weight $\mathbf{w}$ are
decided in the training stage. The out-of-sample performance of our
design methods are tested in the trading stage. %

\begin{figure}[t]
\begin{centering}
\includegraphics[scale=0.5]{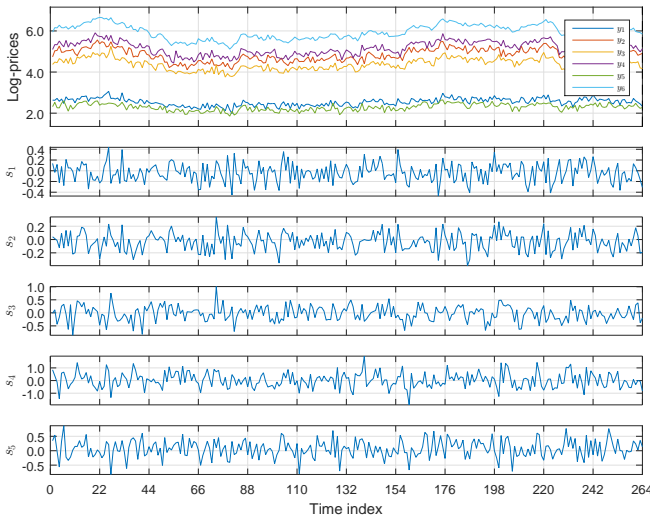}
\par\end{centering}

\protect\caption{\label{fig:Synthetic Trading}Log-prices and five estimated spreads.
(The sample length for in-sample training is chosen to be $5\times12\times22$,
and the sample length for out-of-sample trading is $12\times22$.)}
\end{figure}

In the synthetic experiments, we set $M=6$ and $r=5$ and only show
the performance of the MRP design methods under net budget constraint
${\cal W}_{1}$. We estimate $N=5$ spreads using the generated sample
path. Based on these $5$ spreads, an MRP is designed as $z_{t}=\mathbf{w}^{T}\mathbf{s}_{t}$.
The simulated log-prices and the spreads for the trading stage are
shown in Figure \ref{fig:Synthetic Trading}.

\begin{figure}[t]
\begin{centering}
\includegraphics[scale=0.5]{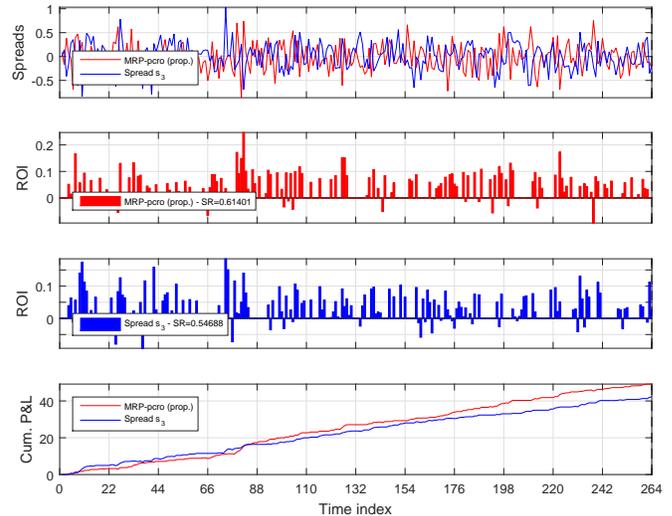}
\par\end{centering}

\protect\caption{\label{fig:Synthetic Performance-1}Comparisons of ROIs, Sharpe ratios
of ROIs, and cumulative P\&Ls between the MRP designed using our proposed
method denoted as MRP-$\text{pre}$ (prop.) with one underlying spread
denoted as Spread $s_{3}$.}
\end{figure}

\begin{figure}[th]
\begin{centering}
\includegraphics[scale=0.5]{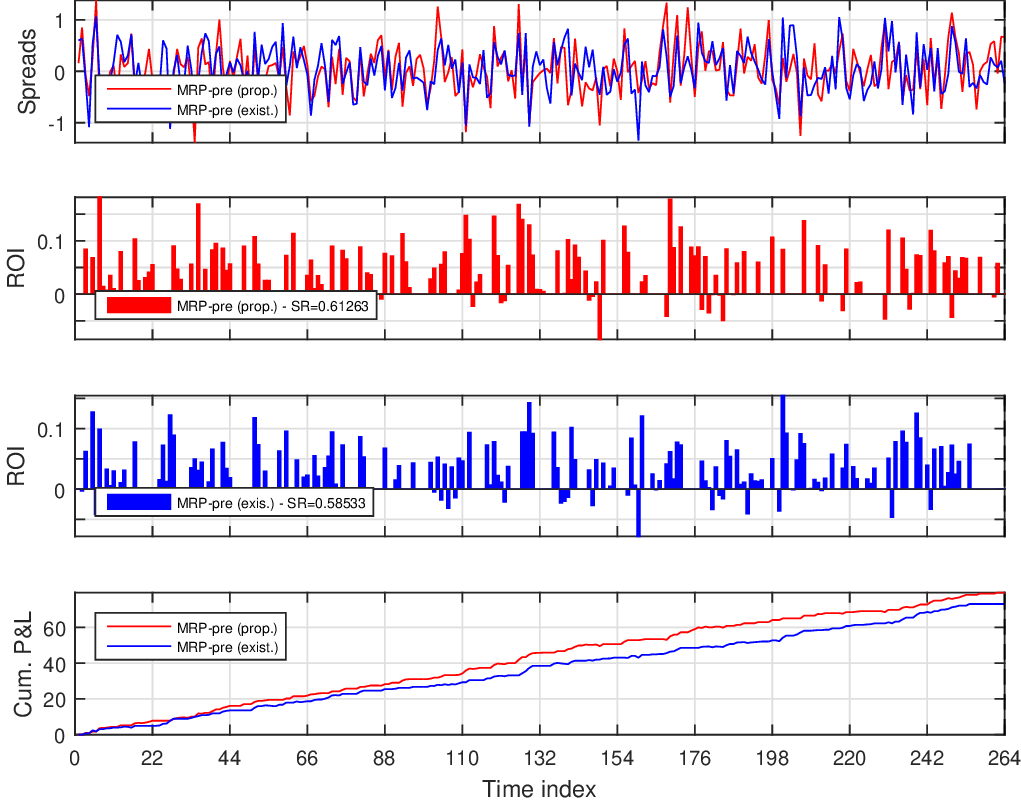}
\par\end{centering}

\protect\caption{\label{fig:Synthetic Performance-2}Comparisons of ROIs, Sharpe ratios
of ROIs, and cumulative P\&Ls between the MRP designed using our proposed
method denoted as MRP-$\text{pcro}$ (prop.) and one existing benchmark
method in \cite{CuturidAspremont2013} denoted as MRP-$\text{pcro}$
(exist.). }
\end{figure}

The performance of the MRP designed using our proposed methods are
compared with those of one underlying spread and the method in \cite{CuturidAspremont2013}
based on $\text{pcro}_{z}\left(5,\mathbf{w}\right)$ and $\text{pre}_{z}\left(\mathbf{w}\right)$,
which are shown in Figure \ref{fig:Synthetic Performance-1} and Figure
\ref{fig:Synthetic Performance-2}. From our simulations, we can conclude
that our designed MRPs do generate consistent positive profits. And
simulation results also show that our designed portfolios can outperform
the underlying spreads and the MRPs designed using methods in \cite{CuturidAspremont2013}
with higher Sharpe ratios of ROIs and higher cumulative P\&Ls. 

\subsection{Market Data Experiments}

\begin{figure}[t]
\begin{centering}
\includegraphics[scale=0.5]{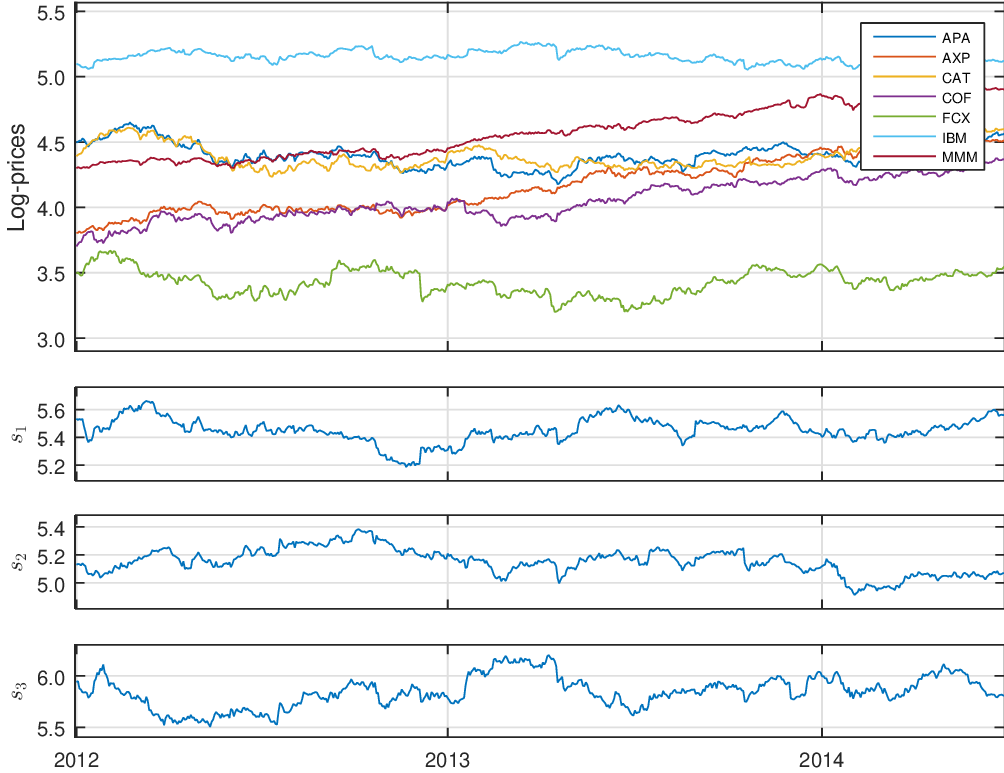}
\par\end{centering}

\protect\caption{\label{fig:real_data_logprices-spreads}Log-prices for $\left\{ \mathsf{APA},\mathsf{AXP},\mathsf{CAT},\mathsf{COF},\mathsf{FCX},\mathsf{IBM},\mathsf{MMM}\right\} $
and three spreads $s_{1}$, $s_{2}$, and $s_{3}$. }
\end{figure}

We also test our methods using real market data from the Standard
\& Poor's 500 (S\&P 500) Index, which is usually considered as one
of the best representatives for the U.S. stock markets. The data are
retrieved from Yahoo! Finance\footnote{\href{http://finance.yahoo.com}{http://finance.yahoo.com}}
and adjusted daily closing stock prices are employed. We first choose
stock candidates which are possibly cointegrated to form stock asset
pools. One stock pool is $\left\{ \mathsf{APA},\mathsf{AXP},\mathsf{CAT},\mathsf{COF},\mathsf{FCX},\mathsf{IBM},\mathsf{MMM}\right\} $,
where the stocks are denoted by their ticker symbols%
. Three spreads are constructed from this pool. Then MRP design methods
are employed and unit-root tests are used to test their tradability.
The log-prices of the stocks and the log-prices for the three spreads
are shown in Figure \ref{fig:real_data_logprices-spreads}. %

\begin{figure}[th]
\begin{centering}
\includegraphics[scale=0.5]{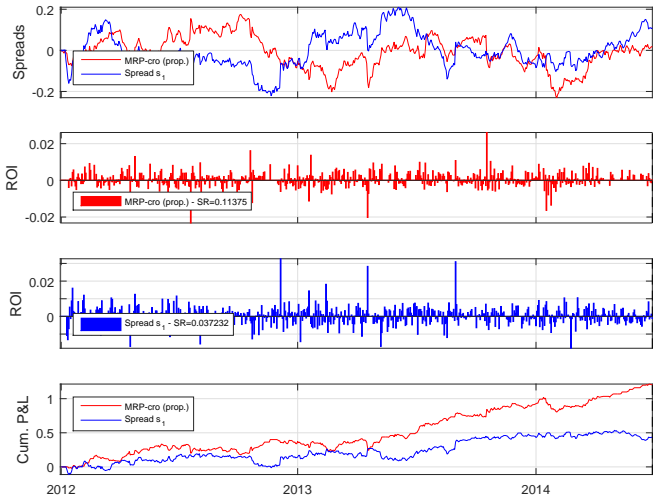}
\par\end{centering}

\protect\caption{\label{fig:real_data_performance-1}Comparisons of ROIs, Sharpe ratios
of ROIs, and cumulative P\&Ls between the MRP designed using our proposed
method denoted as MRP-$\text{cro}$ (prop.) with one underlying spread
denoted as Spread $s_{1}$.}
\end{figure}

Based on the mean-reversion trading framework mentioned before, one
trading experiment is carried out from February 1st, 2012 to June
30th, 2014. In Figure \ref{fig:real_data_performance-1}, we compare
the performance of our designed MRP with the underlying spread $s_{1}$.
The log-prices for the designed spreads, and the out-of-sample performance
like ROIs, Sharpe ratios of ROIs, and cumulative P\&Ls are reported.
It is shown that using our method, the designed MRP can achieve a
higher Sharpe ratio and a better final cumulative return.

\begin{figure}[th]
\begin{centering}
\includegraphics[scale=0.5]{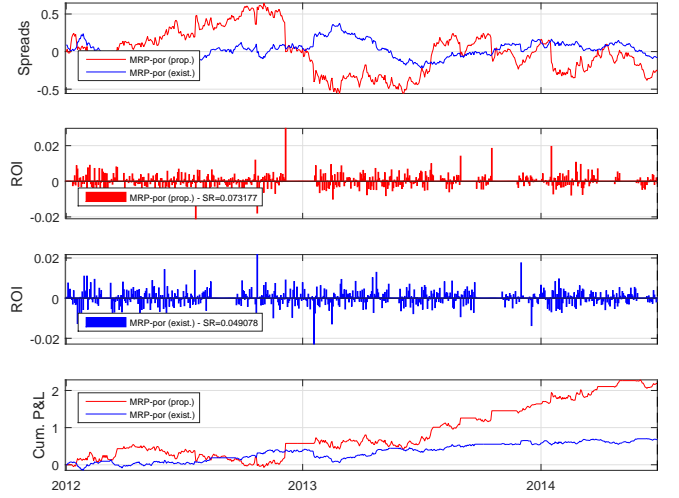}
\par\end{centering}

\protect\caption{\label{fig:real_data_performance-2}Comparisons of ROIs, Sharpe ratios
of ROIs, and cumulative P\&Ls between the MRP designed using our proposed
method denoted as MRP-$\text{por}$ (prop.) and one existing benchmark
method in \cite{CuturidAspremont2013} denoted as MRP-$\text{por}$
(exist.). }
\end{figure}

We also compare our proposed design method with the method in \cite{CuturidAspremont2013}
using $\text{por}_{z}\left(5,\mathbf{w}\right)$ in Figure \ref{fig:real_data_performance-2}.
We can see that our method can outperform the benchmark method in
terms of Sharpe ratio and the return performance. %

\section{Conclusions\label{sec:Conclusions}}

The mean-reverting portfolio design problem arising from statistical
arbitrage has been considered in this paper. We have formulated the
MRP design problem in a general form by optimizing a mean-reversion
criterion characterizing the mean-reversion strength of the portfolio
and, at the same time, taking into consideration the variance of the
portfolio and an investment budget constraint. Several specific optimization
problems have been proposed based on the general design idea. Efficient
algorithms have been derived to solve the design problems. Numerical
results show that our proposed methods are able to generate consistent
positive profits and significantly outperform the the design methods
in literature.

\appendices{}

\section{Proof for Lemma \ref{lem:majorization-overall}\label{Proof-for-Lemma-majorazation-overall}}

For problem \eqref{eq:por and pcro with eq01} with $\mathbf{w}\in{\cal W}_{0}$,
the majorizing function in the first majorization step \eqref{eq:majorization function for por and pcro}
is denoted by $u_{1}\left(\mathbf{w},\mathbf{w}^{\left(k\right)}\right)$,
and the majorizing function in the second step \eqref{eq:majorization function for por and pcro with eq0}
is denoted by $u_{2}\left(\mathbf{w},\mathbf{w}^{\left(k\right)}\right)$.
From the majorization properties in \eqref{eq:majorization property},
then we can have have this relationship: $f_{z}\left(\mathbf{w}\right)\leq u_{1}\left(\mathbf{w},\mathbf{w}^{\left(k\right)}\right)\leq u_{2}\left(\mathbf{w},\mathbf{w}^{\left(k\right)}\right)$. 

Then we can get the overall majorization in $\mathbf{w}$ for the
objective function of the original problem at $\mathbf{w}^{\left(k\right)}$
over the constraint set by the following function:

\[
\begin{array}{cl}
 & u_{2}\left(\mathbf{w},\mathbf{w}^{\left(k\right)}\right)\\
\overset{\left(\text{in }\bar{\mathbf{x}}\right)}{=} & 2\left[\left(\bar{\mathbf{N}}^{\left(k\right)}-\psi\left(\bar{\mathbf{N}}^{\left(k\right)}\right)\mathbf{I}\right)\bar{\mathbf{x}}^{\left(k\right)}\right]^{T}\bar{\mathbf{x}}\\
 & +2\psi\left(\bar{\mathbf{N}}^{\left(k\right)}\right)\nu-\bar{\mathbf{x}}^{\left(k\right)T}\bar{\mathbf{N}}^{\left(k\right)}\bar{\mathbf{x}}^{\left(k\right)}\\
\overset{\left(\text{in }\mathbf{x}\right)}{=} & 2\left[\left(\mathbf{N}^{\left(k\right)}-\psi\left(\mathbf{R}^{-1}\mathbf{N}^{\left(k\right)}\mathbf{R}^{-T}\right)\mathbf{N}_{0}\right)\mathbf{x}^{\left(k\right)}\right]^{T}\mathbf{x}\\
 & +2\psi\left(\mathbf{R}^{-1}\mathbf{N}^{\left(k\right)}\mathbf{R}^{-T}\right)\nu-\mathbf{x}^{\left(k\right)T}\mathbf{N}^{\left(k\right)}\mathbf{x}^{\left(k\right)}\\
\overset{\left(\text{in }\mathbf{w}\right)}{=} & 2\left[\left(\mathbf{H}^{\left(k\right)}-\psi\left(\mathbf{R}^{-1}\mathbf{F}^{T}\mathbf{H}^{\left(k\right)}\mathbf{F}\mathbf{R}^{-T}\right)\mathbf{M}_{0}\right)\mathbf{w}^{\left(k\right)}\right]^{T}\mathbf{w}\\
 & +2\psi\left(\mathbf{R}^{-1}\mathbf{F}^{T}\mathbf{H}^{\left(k\right)}\mathbf{F}\mathbf{R}^{-T}\right)\nu-\mathbf{w}^{\left(k\right)T}\mathbf{H}^{\left(k\right)}\mathbf{w}^{\left(k\right)},
\end{array}
\]
where the last two terms in every step of the derivations are constants
since they are independent of the optimization variables.

\section{Proof for Lemma \ref{lem:spread-return}\label{Proof-for-Lemma-spread-return}}

Since the spread is defined as $z_{t}=\mathbf{w}_{p}^{T}\mathbf{y}_{t}$,
then the multi-period P\&L at time $t$ for $\tau$ holding periods
is given by 

\[
\begin{array}{rl}
\mathrm{P\&L}_{t}\left(\tau\right)= & \mathbf{w}_{p}^{T}\mathbf{r}_{t}\left(t-t_{o}\right)-\mathbf{w}_{p}^{T}\mathbf{r}_{t-\tau}\left(t-\tau-t_{o}\right)\\
= & \sum_{m=1}^{M}w_{p,m}r_{m,t}\left(t-t_{o}\right)\\
 & -\sum_{m=1}^{M}w_{p,m}r_{m,t-\tau}\left(t-\tau-t_{o}\right)\\
= & \sum_{m=1}^{M}w_{p,m}\left(\frac{p_{m,t}}{p_{m,t_{o}}}-1\right)\\
 & -\sum_{m=1}^{M}w_{p,m}\left(\frac{p_{m,t-\tau}}{p_{m,t_{o}}}-1\right)\\
\approx & \sum_{m=1}^{M}w_{p,m}\left[\log\left(p_{m,t}\right)-\log\left(p_{m,t_{o}}\right)\right]\\
 & -\sum_{m=1}^{M}w_{p,m}\left[\log\left(p_{m,t-\tau}\right)-\log\left(p_{m,t_{o}}\right)\right]\\
= & \sum_{m=1}^{M}w_{p,m}\log\left(p_{m,t}\right)-w_{p,m}\log\left(p_{m,t-\tau}\right)\\
= & \sum_{m=1}^{M}w_{p,m}y_{m,t}-\sum_{m=1}^{N}w_{p,m}y_{m,t-\tau}\\
= & \mathbf{w}_{p}^{T}\mathbf{y}_{t}-\mathbf{w}_{p}^{T}\mathbf{y}_{t-\tau}\\
= & z_{t}-z_{t-\tau}.
\end{array}
\]
The approximation in the fourth step follows from $\log\left(1+x\right)\approx x$
when $x\rightarrow0$, where $\log\left(\cdot\right)$ denotes the
natural logarithm. Similarly, for a short position on the MRP, the
calculation of $\mathrm{P\&L}_{t}\left(\tau\right)$ is given by $z_{t-\tau}-z_{t}$.

\bibliographystyle{IEEEtran}
\bibliography{MRP}

\end{document}